\documentclass[fleqn,usenatbib]{mnras}

\usepackage[T1]{fontenc}

\DeclareRobustCommand{\VAN}[3]{#2}
\let\VANthebibliography\thebibliography
\def\thebibliography{\DeclareRobustCommand{\VAN}[3]{##3}\VANthebibliography}

\usepackage{graphicx}	
\usepackage{amsmath}	
\usepackage{amssymb}	
\usepackage{wasysym}
\usepackage{caption}
\usepackage{subcaption}
\usepackage{hyperref}
\usepackage{iondefs}
\usepackage{threeparttable}
\usepackage{newtxtext,newtxmath}
\usepackage[dvispnames]{xcolor}
\usepackage{deluxetable}
\usepackage{longtable}

\definecolor{my_red}{RGB}{255, 112, 116}
\definecolor{my_purple}{RGB}{143, 120, 174}
\definecolor{my_orange}{RGB}{231, 126, 61}
\definecolor{my_green}{RGB}{138, 201, 38}
\definecolor{my_blue}{RGB}{25, 130, 196}
\definecolor{my_yellow}{RGB}{255, 202, 58}
\definecolor{my_g1_green}{RGB}{100, 167, 71}
\definecolor{my_g1_blue}{RGB}{66, 103, 172}
\definecolor{my_pink}{RGB}{199, 105, 134}

\title[Complex CGM of Galaxy Groups]{Unveiling the Complex Circumgalactic Medium: A Comparative Study of Merging and Non-Interacting Galaxy Groups}

\author[A.~Fern\'{a}ndez-Figueroa et al.]
{Antonia Fern\'{a}ndez-Figueroa$^{1,2}$\thanks{E-mail: antoniafernandez@swin.edu.au},
Glenn G.~Kacprzak,$^{1,2}$ Nikole M.~Nielsen$^{1,2}$, Tania M. Barone$^{1,2,3}$,
\newauthor{Hasti Nateghi$^{1,2}$, Sameer$^{4}$, Deanne B.~Fisher$^{1,2}$ and Bronwyn Reichardt Chu$^{1,2,5,6}$}
\\
$^{1}$Centre for Astrophysics and Supercomputing, Swinburne University of Technology, Hawthorn, Victoria 3122, Australia\\
$^{2}$ARC Centre of Excellence for All Sky Astrophysics in 3 Dimensions (ASTRO 3D), Australia\\
$^{3}$School of Physics, University of New South Wales, Kensington, Australia\\
$^{4}$Department of Physics and Astronomy, The University of Notre Dame\\
$^{5}$Centre for Extragalactic Astronomy, Department of Physics, Durham University, South Road, Durham DH1 3LE, UK\\
$^{6}$Institute for Computational Cosmology, Department of Physics, Durham University, South Road, Durham DH1 3LE, UK\\
}

\date{Accepted 2024 May 17. Received 2024 May 17; in original form 2024 March 14}

\pubyear{2024}

\begin{document}
\label{firstpage}
\pagerange{\pageref{firstpage}--\pageref{lastpage}}
\maketitle

\begin{abstract}

While most galaxies live in group environments where they undergo an accelerated evolution, the characteristics of their circumgalactic medium (CGM) remain uncertain. We present an analysis of the CGM of two galaxy groups in different stages of interaction: (G1) a close pair of galaxies ($z=0.043$) separated by 87~kpc that do not show signs of interactions and (G2) four merging galaxies ($z=0.098$) separated by 10~kpc. We present spatially-resolved Keck/KCWI galaxy observations and {\it HST}/COS quasar spectra (G1 at 48~kpc and G2 at 100~kpc away) to quantify both the resolved galaxy and CGM properties in these two different group environments. G1 contains two typical star-forming galaxies with no evidence of strong outflows. G2 contains two star-forming, one post-starburst and one quiescent galaxy. Both groups have a range of CGM detected metal lines ({\HI},~{\CII},~{\SiII},~{\SiIII},~{\NV}~and~{\OVI}). Despite G2 being twice as far from the quasar, G2 has $\log (N({\HI})/{\rm cm}^{-2})=17.33$, compared to $\log (N({\HI})/{\rm cm}^{-2})=16.43$ for G1. We find that the CGM of the merging galaxies (G2) is more kinematically complex, is in a higher ionisation state, spans a wider range of metallicities and column densities, has smaller cloud sizes, and is inconsistent with the simple superposition model that seems to match well with G1. We conclude that the complexity of the CGM in merging galaxies surpasses that of not strongly interacting galaxies, suggesting that mergers play a significant role in shaping the intricate structure of the CGM.
\end{abstract}




\section{Introduction}

The gas that resides outside the interstellar medium but within the virial radius of galaxies is known as the circumgalactic medium (CGM), where approximately 50\% of the baryonic mass in galaxy halos are located \citep{Werk2014}. Various processes associated with the CGM such as outflows, inflows and gas recycling are fundamental to understanding galaxy evolution  \citep[e.g.,][]{Tumlinson_2017}. The CGM is normally too diffuse to detect in emission, so it is typically detected through absorption, using bright background quasars \citep[QSOs; e.g.][]{Lanzetta_1995, Chen_1998a, Churchill_2000a, Kacprzak2008, Nielsen_2013b, Prochaska_2017b, Peroux2022} or gravitationally lensed galaxies \citep[][]{Lopez_2018, Lopez_2020c, Mortensen2021, Tejos2021, Fernandez-Figueroa2022, Barone2024} as background sources. 

Most galaxies live in group environments or were located in one in the past \citep[e.g.][]{Tully_1987, Garcia_1993, Eke_2004c}. Therefore, understanding the group CGM is fundamental to understanding the evolution of most galaxies in the Universe. However, the vast majority of CGM studies have not considered galactic environment, leaving many unanswered questions about a vital period in every galaxy's history. Recent work has found that the CGM in group environments is more extended and kinematically complex \citep{Chynoweth_2008, Chen_2010a, Bordoloi_2011a, deBlok_2018, Nielsen_2018b, Dutta2020, Huang2021}, going to distances up to 140~kpc from the centre of the group, although it is anisotropic in its spatial distribution \citep{Fossati2019, McCabe_2021b}. Moreover, it has been theorised that the CGM of groups is directly affected by interactions between their galaxies, such as mergers, tidal streams and gas transfers between galaxies, making the CGM more complex in these cases \citep{Bowen_1995, ChurchillandCharlton_1999, Kacprzak_2010a, Kacprzak_2010b, AnglesAlcazar2017, Nateghi2021, Nielsen2022}. This opens the possibility that the CGM absorption associated with group galaxies belongs to multiple galaxies within the group \citep{Kacprzak_2010b, Bielby_2017, Peroux_2017, Peroux_2019, Pointon_2017a, Pointon_2020, Rahmani_2018, Nielsen_2018b, Hamanowicz_2020, Lehner_2020, Cherrey2023}. 

While \MgII~absorption strength tends to be higher in group environments \citep{Bordoloi_2011a, Nielsen_2018b, Dutta2018}, multiple observational studies have found a decrease in both \CIV~\citep{Burchett2016} and \OVI~\citep{Pointon_2017a, McCabe_2021b} covering fractions and kinematics associated with group galaxies at low redshifts. This trend is due to the CGM of group galaxies having a higher temperature \citep{Bielby_2017, Ng2019, Zahedy2019} that allows ions to go into higher ionisation states, a phenomenon also found in simulations \citep{Oppenheimer2016, Oppenheimer2021, Wijers2022}. It is important to note that while the CGM of group galaxies is dominated by the hot phase ($T\sim10^7$~K), there is still gas in the warm ($T\sim10^5$~K) and cool ($T\sim10^4$~K) phases.

Galaxy mergers are quite common in group environments. It is well-known that interactions between galaxies trigger star formation both from simulations \citep{BarnesandHernquist_1996, MihosandHernquist_1996, Tissera_2002, DiMateo_2007, Montuori_2010, Renaud_2014} and observations \citep{LarsonandTinsley_1978, BartonGellerandKenyon_2000, Alonso_2004, NikolicCullenandAlexander_2004, WoodsGellerandBarton_2006, WoodsandGeller_2007, Ellison_2008, Heiderman_2009, KnapenandJames_2009, Robaina_2009, Ellison_2010, Patton_2011, Pan_2018}. Furthermore, star formation drives outflows \citep[e.g.][]{Heckman2000, Rubin_2014, ReichardtChu2022, ReichardtChu2024}, which could indicate that the CGM of merging galaxies might be dominated by outflowing gas. Indeed, simulations suggest this is true, \citet{Hani_2018} found that mergers increase the physical extent and metallicity of the CGM, and that outflowing gas dominates the CGM kinematics. \citet{Sparre2022} found that merging events can decrease the metallicity in the centre of the host galaxy through these outflows. Simulations further suggest that one of the causes of the increased temperature in the group CGM is heating through mergers \citep{Cox2004, Cox2006b, Sinha2009, Moreno2019}. However, a later study concluded that mergers are not the main determinant of the CGM in group environments, with other processes, such as AGN feedback, having a more important role \citep{Hani2019}. Observational studies have found that merging events can be associated with higher \HI~column densities \citep{Dutta2018}, large velocity offsets between absorbers and their host galaxies \citep{Keeney2011, Johnson2015, Rupke2019}, and diluted metallicities in their host galaxies \citep{Bustamante2020, Calabro2022}. Unfortunately, observational studies of the CGM of merging galaxies have focused on small samples and inhomogeneous environments, so more studies are necessary to understand its characteristics.

\begin{figure}

    \begin{subfigure}[b]{\columnwidth}
    \centering
    \includegraphics[width=\columnwidth]{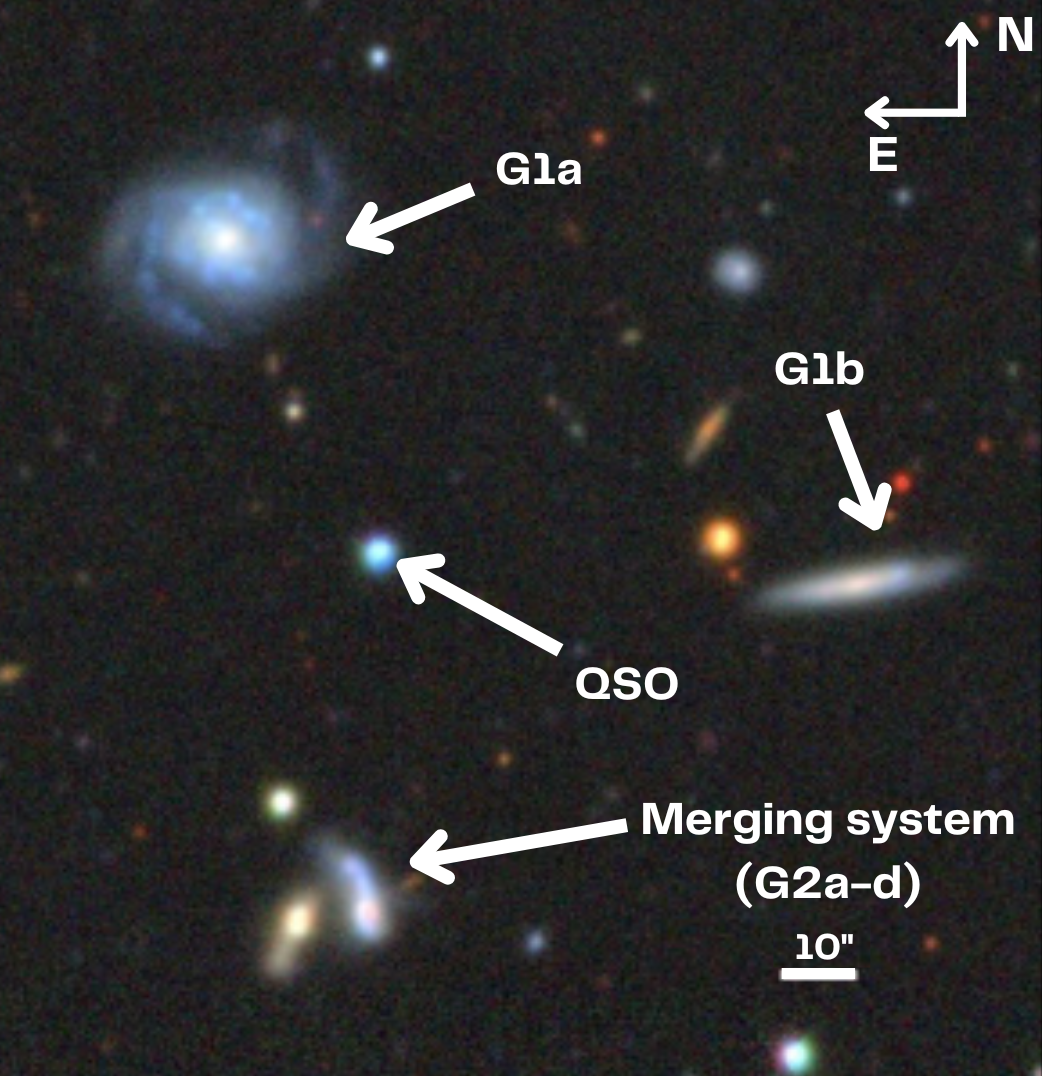}
    \end{subfigure}

    \begin{subfigure}[b]{\columnwidth}
    \centering
    \includegraphics[width=\columnwidth]{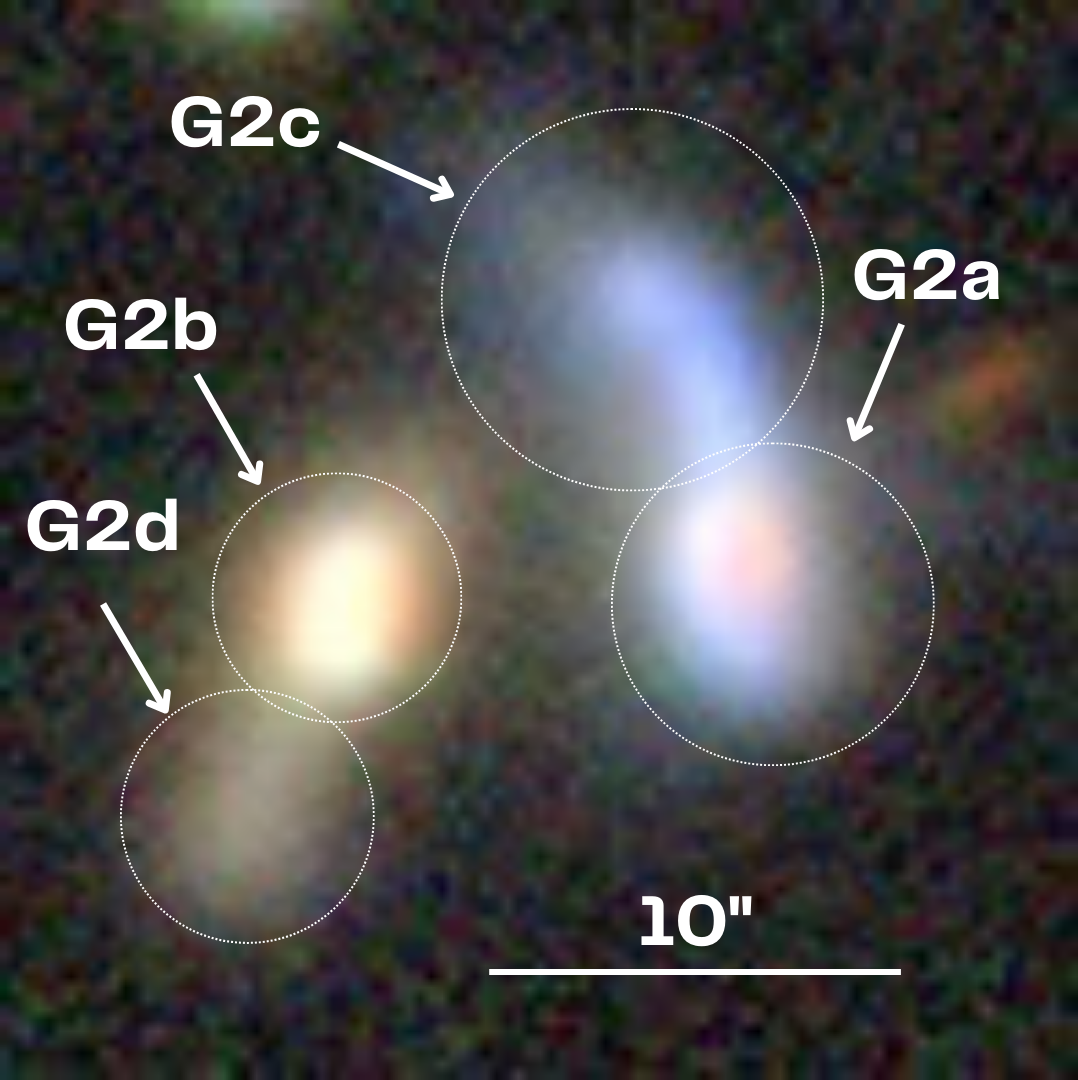}
    \end{subfigure}

   \caption{\textit{Top:} DECaLS $grz$ image of the quasar field with two intervening galaxy groups. At the top we have G1a and G1b, a pair of galaxies at $z=0.043$ which are not strongly interacting, the background QSO is at the center and G2a-d, the strongly interacting group of four merging galaxies at $z=0.098$, is at the bottom. There is an unrelated object to the North-East of the merging system. \textit{Bottom:} zoom-in into the merging system, highlighting each of the galaxies that belong to it and the labels that will be used throughout this paper.}
    \label{fig:system}
\end{figure}

To further investigate the effect of mergers in the CGM, we analyse two galaxy groups located near the line of sight of the same QSO. The first group, hereafter Group 1, features two nearby galaxies at $z=0.043$ that are separated by 87~kpc and do not appear to be strongly interacting (see Fig.~\ref{fig:system} \textit{(top)}). The QSO line-of-sight is located 40~kpc from G1a and 55~kpc from G1b. The second group, hereafter Group 2, has four merging galaxies. The configuration of this group is displayed in Fig.~\ref{fig:system} \textit{(bottom)}. It contains three star-forming galaxies (G2a-c) and one quiescent galaxy (G2d), all located at $z_{\rm G2}=0.09849$. The QSO line-of-sight is located 100~kpc from the galaxies. These two systems have galaxies at different stages of interaction. As such, we may expect to find key differences in their CGM, which we explore in this paper. Using Keck/KCWI IFU data, we are able to directly compare the resolved galaxy properties to those of their CGM. 

This paper is organised as follows: Section \ref{data_and_analysis} details our datasets and all the methods used to analyse them. Results of the distant pair are described in Section \ref{distant_pair_results}, while those of the merging system are detailed in Section \ref{merging_results}. These results are consequently discussed in Section \ref{discussion}. Finally, Section \ref{conclusions} offers a summary and presents the final conclusions of the paper. We assume a flat $\Lambda$ cold dark matter ($\Lambda$CDM) Universe with $\Omega_{\Lambda}= 0.7$, $\Omega_{M} = 0.3$, and $H_{0} = 70$~km~s$^{-1}$~Mpc$^{-1}$, and a Kroupa initial mass function \citep{Kroupa_2001b}.

\begin{table*}
	\centering
	\caption{Properties of the galaxies.}
	\label{tab:morphology}
	\begin{tabular}{lccccccccc}
     \hline
        Galaxy & RA & DEC & $z$ & $D$ & $i$ & B/T & $\Phi$ & SFR & $A_{v}$ \\
         &  &  &  & (kpc) & ($^\circ$) &  & ($^\circ)$ & (M$_{\astrosun}$~yr$^{-1}$) &  \\
        & (1) & (2) & (3) & (4) & (5) & (6) & (7) & (8) & (9) \\
	\hline
        G1a & 08:32:21.60 & +24:31:42.61 & 0.04320 & 40.4 & 40.30$\substack{+0.10 \\ -0.05}$ & 0.16$\substack{+0.00 \\ -0.00}$ & 67.90$\substack{+0.08 \\ -0.08}$ & 3.68$\pm$0.10 & 0.75 \\
        G1b & 08:32:15.34 & +24:30:56.27 & 0.04308 & 55.5 & 89.00$\substack{+1.00 \\ -0.03}$ & 0.00$\substack{+0.00 \\ -0.00}$ & 13.59$\substack{+0.20 \\ -0.24}$ & 0.54$\pm$0.02 & 0.62  \\
        G2a & 08:32:20.23 & +24:30:11.45 & 0.09840 & 97.7 & - & - & - & 5.13$\pm$0.7 & 0.68 \\
        G2b & 08:32:20.89 & +24:30:10.61 & 0.09875 & 102.2 & - & - & - & 0.06$\pm$0.03$^{a}$ & 0.75 \\
        G2c & 08:32:20.26 & +24:30:16.22 & 0.09897 & 89.1 & - & - & - & 2.30$\pm$0.0 & -$^{c}$ \\
        G2d & 08:32:21.03 & +24:30:06.51 & 0.09782 & 110.0 & - & - & - & $\leq 0.03^{b}$ & -$^{c}$ \\
        \hline
	\end{tabular}
 
\raggedright The columns are: (1) Right ascension, (2) declination, (3) redshift, (4) impact parameter, (5) inclination angle, (6) bulge-to-total ratio, (7) azimuthal angle,(8) total star-formation rate, and (9) mean $A_{v}$ parameter. $^{a}$\footnotesize{The SFR was calculated using \OII.} $^{b}$\footnotesize{It was not possible to calculate SFR because this galaxy has no emission lines (see Section \ref{SFR})}. $^{c}$\footnotesize{It was not possible to calculate $A_{v}$ because these galaxy have no Hydrogen emission lines.}
\end{table*}

\section{Data and analysis} \label{data_and_analysis}

The quasar field we studied in this paper is displayed in Fig.~\ref{fig:system} and properties of the galaxies are shown in Table~\ref{tab:morphology}. We have observed two galaxy groups in the quasar field J0832$+$243.

The groups were originally identified as a single candidate compact group by \citet{McConnachie2009}, given the isolated nature of four bright galaxies in proximity to each other. Upon spectroscopic follow-up, the visually identified compact group was actually two groups. There are no other objects that have been identified at the redshift of our now two groups. However, the Arecibo Legacy Fast ALFA survey \citep[ALFALFA; ][]{Haynes2011}{}{} has identified \HI~detection at the redshift of G1. This provides a \HI~mass of $\log({\rm M/M_{\astrosun}})=10.1$.

In the following subsections, we present the data reduction and analysis of our Keck/KCWI IFU data, {\it HST}/COS spectroscopy, and Sloan imaging of the field. The IFU data were used to obtain resolved galaxy properties, the COS data were used to characterize the physical conditions of the multiphase CGM, and the Sloan imaging was used to model the morphological properties of the galaxies.

\subsection{KCWI integral field spectroscopy} \label{KCWI}

Observations of the galaxies were performed with the Keck Cosmic Web Imager \citep[KCWI;][]{Morrissey2018} on 2022 March 3 UT (PID: 2022A\_W230) using the large slicer and BL grating with a wavelength centre of 4500~\AA~and $2\times2$ binning. The large slicer has a field-of-view of $33''\times20.4''$. The BL grating has a resolving power of 900 (1.75~{\AA}~pix$^{-1}$ or 128~km~s$^{-1}$) and a wavelength range of $3500 \lesssim \lambda \lesssim 5500$~{\AA}. We took one exposure centred on G1a, another centred on G1b and a third centred on the G2 group. Each of these exposures was 300~s long. 

We used the IDL version of the KCWI Data Reduction Pipeline (DRP)\footnote{\url{ https://github.com/Keck-DataReductionPipelines/KcwiDRP}} with standard settings but skipped the sky subtraction step to aid in additional flat-fielding at a later stage. A flux calibration was performed in the last step of the DRP using the standard star bd33d2642 from the DRP star list. A residual illumination gradient was later removed and the sky was subtracted following the procedure of \citet{Nielsen2022}. The data were reprojected onto square spaxels with $0.29''$ sides using {\sc Montage}\footnote{\url{http://montage.ipac.caltech.edu/}} with default settings. Because this reprojection oversamples the original data (we did not dither the observations since we only obtained one exposure per galaxy), these spaxels were then spatially binned $3\times3$, resulting in a pixel size of $0.87\times0.87$~arcsec$^2$ to match the seeing of the observations. While this binning accounts for most of the oversampling, the final rebinned cube is oversampled in the slice direction by 1.6 spaxels. The final spatial resolution of each spaxel at the redshift of the galaxies is 0.72, 0.69 and 1.08 kpc for the G1a, G1b and G2 respectively. Additional details about the reduction process can be found in \citet{Nielsen2022}. The Galactic extinction was corrected using a \citet{Cardelli_1989} attenuation law. The value of the $A_{V}$ parameter at the right ascension and declination of the galaxies was calculated using the NED Extinction Calculator.\footnote{\url{https://ned.ipac.caltech.edu/extinction_calculator}} 

The pPXF software \citep{Capellari_2004, Capellari_2017a, Capellari_2022}, which is designed for stellar continuum modelling, was used to fit the stellar {\Hbeta} absorption in order to obtain the correct {\Hbeta} emission-line flux. We used the templates from the Extended Medium resolution INT Library of Empirical Spectra \citep[E-MILES,][]{Sanchez_Blazquez_2006b}, with BaSTI isochrones \citep{Pietrinferni2004, Pietrinferni2006}, and a Kroupa IMF. These templates span a range of 53 ages from 0.03 to 14.00 Gyr, 12 [Z/H] values from $-2.27$ to $0.40$, and have [$\alpha$/Fe] values scaled to the solar neighbourhood \citep[base models;][]{Vazdekis2016}. The software was run with an additive polynomial of order 12 \citep[see][]{VanDeSande2017}. The pPXF continuum subtraction was performed only on the spaxels that had a continuum ${\rm SNR}>5$, along with an additional polynomial fit around the emission lines we were interested in ({\OIIedblt}, {\OIIIedblt}, {\Hbeta} and {\Hgamma}). This second polynomial fit was used to anchor the zero-flux level while fitting Gaussians to the emission lines. In the case of spaxels with ${\rm SNR}<5$, only the polynomial fit around the lines was performed.

After subtracting the continuum, we corrected for the intrinsic dust extinction within each galaxy using the equations described in \citet{Cardelli_1989}. The $A_{V}$ parameter was calculated using the ratio between the flux of the {\Hbeta} and the {\Hgamma} lines. Equations 3 and 4, and Table 2 of \citet{Calzetti_2001} were used to calculate $E(B-V)$. The flux from the two lines was calculated by fitting a Gaussian model to them. Both lines were assumed to come from the same gas so that they would have the same velocity and velocity dispersion. Having calculated $E(B-V)$, we transformed it into $A_{V}$ assuming the standard of $R_{V}=3.1$. The average $A_{V}$ values for each galaxy are presented in Table~\ref{tab:morphology}. 

\subsection{Galaxy physical properties}

\subsubsection{Galaxy morphologies}

SDSS r-band images were used to model galaxy properties such as inclinations, azimuthal angles and impact parameters \citep[e.g.,][]{Kacprzak2011, Kacprzak2012, Kacprzak2019} by fitting a bulge+disk model to the galaxies in the G1 pair using Galaxy IMage 2D \citep{Simard_2002a} where the models were convolved with the instrument's point spread function. The surface brightness of the disk component was modelled using an exponential profile, while the bulge component was modelled using a Sersic profile \citep{Sersic_1968}. The galaxies' centroids were extracted using Source Extractor \citep{Bertin_1996}. It was not possible to perform this analysis on G2 galaxies, as their morphology is more complicated. This work defines azimuthal angle ($\Phi$) as the angle between the galaxy's projected major axis and the background quasar sightline. Results of this analysis are shown in Table~\ref{tab:morphology}.

\subsubsection{Gas-phase kinematics} \label{kinematics}

The spectra of G1a, G1b, G2a and G2b have significant detections of the {\OIIedblt} ($\geq 5\sigma$), {\Hbeta} ($\geq 3\sigma$) and {\OIIIe}~$\lambda 5007$ ($\geq 3\sigma$) emission lines. G2c only has significant {\OIIe} ($\geq 5\sigma$) emission lines detected. Using an in-house Python software, we simultaneously fitted a single Gaussian to the {\Hbeta} and {\OIIIe} lines and a double Gaussian to the {\OIIedblt} doublet, assuming that all the lines for a given galaxy have the same velocity centroid. Additionally, we assumed the two lines in the {\OIIedblt} doublet have the same FWHM and that the ratio in the amplitude of the two lines range between 0.5 and 2. The results of these fits were used to create velocity maps and measure the physical properties of the galaxies. G2d did not have any emission lines and we are unable to compute gas-phase properties for this galaxy. The stellar kinematics for G2d are described in Section 2.2.5.

The galaxies in G2 were deblended by analysing the kinematics of the system. We found that G2c is rotating perpendicular to G2a, while G2d has a much lower redshift compared to G2b. Additionally, we looked into differences in the spectra of the different galaxies, with the difference between G2b and G2d being the most significant. G2b has the characteristics of being a post-starburst galaxy (its \Hdelta~equivalent width is 4.3 \AA), while G2d does not have any emission lines and only Ca~\textsc{ii}~H\&K absorption is present, indicating that we are looking at different galaxies. 

 Here, we modelled the emission (G1a, G1b, G2a, G2b, G2c) and absorption (G2d) kinematics of our galaxies using the 3DBarolo software \citep{DiTeodoro2015}, which creates a 3D rotating disk model to the kinematic data by fitting a rotation model in concentric rings. A first fit was performed for each of our galaxies, allowing us to derive a rotation curve for each of them. The rotation curve was assumed to have the following shape:
\begin{equation}
    v(r) = v_{\rm max} \tanh{\frac{r}{r_{v}}},
\end{equation}
where $v_{\rm max}$ is the maximum velocity and $r_{v}$ is the turnover radius. Best-fit parameter values for each of our galaxies were found, which allowed us to extend the models up to the location of the quasar. The software was used a second time with the extended parameters and an extended rotation model was obtained to compare with the quasar absorption kinematics.

\subsubsection{Star Formation Rates} \label{SFR}

The star formation rates (SFR) were calculated using the following equation:
\begin{equation}
    {\rm SFR} = C_{\tiny \Halpha} \frac{L_{\tiny \Halpha}}{L_{\tiny \Hbeta}} 10^{-0.4 A_{\tiny \Hbeta}} L'_{\tiny \Hbeta},
\end{equation}
where ${L_{\tiny \Halpha}}/{L_{\tiny \Hbeta}}$ is the intrinsic luminosity ratio \citep{Calzetti_2001}, $L'_{\tiny \Hbeta}$ is the observed {\Hbeta} luminosity, and $C_{\tiny \Halpha}=10^{-41.257}$ is the scale parameter from \citet{Hao_2011b}, assuming a Kroupa IMF \citep{Kroupa_2001b}, $Z_{\astrosun}$, 100 Myr model. The extinction parameter $A_{\tiny \Hbeta}$ was assumed to be 0 since the dust extinction correction had already been performed. 
While it is covered, we are not able to detect {\Hbeta} emission in G2c, so its SFR has to be calculated using the {\OIIe} emission lines. The method described in \citet{Kewley_2004d} was used. This method assumes a Salpeter IMF \citep{Salpeter_1995}, so we changed to a Kroupa IMF using the conversion described in \citet{Speagle_2014b}. G2d does not have any emission lines, so we could only compute the SFR 3$\sigma$ upper limits from the {\Hbeta} line.

Outflows are found to be more common at higher $\Sigma_{\rm SFR}$, and observations of them become quite rare below $\Sigma_{\rm SFR}$~<~0.1~M$_{\astrosun}$~yr$^{-1}$~kpc$^{-2}$ \citep{Heckman2002, Heckman2015, ReichardtChu2022}. We, therefore, derive $\Sigma_{\rm SFR}$ measurements using the aforementioned SFR and the sizes of our spaxels measured in Section~\ref{KCWI}.

\subsubsection{Gas-phase metallicities and ionization}

The metallicity of the gas in the CGM can be used as an indicator of its origin. When the CGM gas has a similar or higher metallicity compared to its host galaxy it is likely coming from outflows \citep[e.g.,][]{Peroux2020, Cameron2021}. Conversely, if the gas has a lower metallicity compared with its host galaxy, it might be attributed to tidal streams and, for the lowest metallicities, accretion. Here, the $R_{23}$ and $O_{32}$ parameters were used to estimate oxygen abundances and ionisation parameters of the galaxies' ISM using the methods of \citet{Kobulnicky_2004a}. A first initial value for the oxygen abundance was calculated using the methods from \citet{Zaritsky_1994} and \citet{McGaugh_1991} and averaging the results. This initial value was used to decide which branch of $R_{23}$ to use. In this method, oxygen abundances and ionisation parameters are highly dependent on each other, so it is necessary to iterate until the method converges to a valid solution for both quantities. Only two iterations were needed in this case before the method converged. These values and maps will be used to compare the ISM with our CGM measurements. In order to compare the CGM to the galaxy ISM, we normalise the ISM metallicity to ${{\rm Z}_{\astrosun}\approx0.015}$ and ${12+\log({\rm O/H})_{\astrosun}=8.72}$ \citep{AllendePrieto2001}.

We also modelled the metallicity gradients of our galaxies. Following the methods of \citet{Klimenko2023}, we compute two gradients: 1) an azimuthally averaged gradient using all the possible spaxels and 2) one assuming an angular window of $30^\circ$. Both are computed since there could be azimuthally-dependent variations in the metallicity gradients. Both are fit using a first-order polynomial. It is expected that earlier-type galaxies have negative metallicity gradients, while the slope becomes more shallow in the case of late-type galaxies \citep{Sanchez2020}.

\subsubsection{Stellar population modelling}

We used pPXF to model the kinematics of the stellar populations of galaxies G1a, G1b, G2a, G2b, and G2c. The E-MILES templates were once again used with BaSTI isochrones and a Kroupa IMF. The emission lines were masked to keep only the stellar component of the spectra and corrected for Galactic extinction and intrinsic dust extinction. We used spaxels with a continuum SNR of $\geq 5\sigma$ for the analysis. An order 12 additive polynomial was used to model the stellar kinematics.

In the case of G2d, the SNR was too low to perform reliable modelling with pPXF, but its spectra featured Ca~\textsc{ii}~H\&K absorption. To study the kinematics of this galaxy, further binning was applied to our KCWI spectra and a double Gaussian model was fitted to the absorption. The results of these fits were used to model the kinematics of this galaxy following the same methods explained in Section~\ref{kinematics}.

\subsection{Quasar spectroscopy and photoionization modelling}

The CGM analysis is based on {\it HST}/COS observations of the QSO J083220+243100 ($z_{\rm qso}=1.30325$) obtained under the Program ID 14071 (PI Borthakur). The COS observations at intermediate resolution (R$\sim$20,000) span the wavelength range $1140 - 1450$ {\AA} with total exposure of 4800 s in the G130M far-UV grating. The coadded spectrum was taken from the HSLA archive \citep{Peeples2017}. The spectrum was rebinned to two wavelength pixels per resolution element of 0.06 {\AA}. The rebinned spectrum was continuum normalized using lower-order polynomials to define the continuum level. The two galaxy groups are located at different redshifts ($z=0.043$ for the G1 group and $z=0.098$ for the G2 group), so their absorption lines do not overlap. At these redshifts, we have spectral coverage for {\Lya}, {\CII}, {\SiII}, {\SiIII}, {\NV}, and {\OVI}. However, we do note that cloud \textcolor{my_purple}{2} in G2 (see Fig.~\ref{fig:qso_abs_G2} and Tables~\ref{tab:abs_properties_G2} and~\ref{tab:vp_properties_G2}) may be blended with a tentative absorption system that could contain {\Lyb} and {\CIII} at only $z = 0.3571$. Thus, values determined for cloud \textcolor{my_purple}{2} should be taken with caution.

Group environments are likely to have a complicated CGM from galaxy--galaxy interactions and there are usually multiple velocity components, or clouds, along a given line of sight. We modelled the absorption in both systems using the Cloud-by-cloud Multiphase Bayesian Modelling method described in \citet{Sameer_2021b, Sameer_2022d, Sameer2024} to quantify those variations. This method uses \textsc{Cloudy} \citep{Ferland_2017} to characterize the physical conditions of the different clouds that produce the absorption in the data. \textsc{Cloudy} is used to generate a grid with different values of metallicity ($Z$), hydrogen number density ($n_{\rm H}$), and neutral hydrogen column density ($N(\HI)$), at the redshift of the absorption, assuming the KS19~\citep{Khaire2019} extragalactic background radiation as the photoionizing radiation field. This grid is then explored by adopting a nested-sampling approach implemented using PyMultinest~\citep{Buchner2014}. Depending on the number of clouds needed to explain the observed absorption in multiple transitions, Voigt Profiles (VPs) are synthesized for each cloud (convolved with the line spread function for the appropriate COS grating) and compared with the observed data in multiple transitions, yielding a distribution for modelled parameters. Typically, all of the observed line transitions cannot be reproduced with one phase of gas. In many instances, a photoionization model that accurately reproduces the observed low-ionization absorption (e.g., {\SiII}, {\SiIII}) fails to generate requisite high-ionization absorption (e.g., {\NV}, {\OVI}). Under such circumstances, we model such absorption using hybrid models with contributions from collisional and photoionization. After applying this method, each cloud in the absorption system is characterized by six parameters: $\log Z$, $\log n_{\rm H}$, $\log T$, $\log N(\rm HI)$, $z$, and $b_{nt}$. We adopt uniform priors on $\log Z$, $\log n_{\rm H}$, $\log T$, and $\log N(\rm HI)$ in the ranges $\log Z$ $\in$ [$-3.0$,1.5], $\log n_{\rm H}$ $\in$ [$-6.0$, 0.0], $\log T$ $\in$ \hbox{[2.0, 6.5]}, $\log N(\rm HI)$ $\in$ [12.0, 19.0]. We use a uniform prior on $b_{nt}$ (the non-thermal contribution to the Doppler parameter of a cloud) ranging between [0, $b+2\sigma(b)$] where $b$ is the Doppler broadening parameter determined from a preliminary VP fit. We determine the Doppler broadening parameter $b$ for all transitions using the equation $b^{2}=b^{2}_{\mathrm{nt}}+b^{2}_{\mathrm{t}}$, where $b_{\mathrm{t}}=\sqrt{2kT/m}$ is the line broadening due to temperature and $b_{\mathrm{nt}}$ the line broadening due to non-thermal effects. The non-thermal broadening component is assumed to be the same for all the transitions in the same cloud. 

\begin{figure*}
    \centering
    \includegraphics[width=\textwidth]{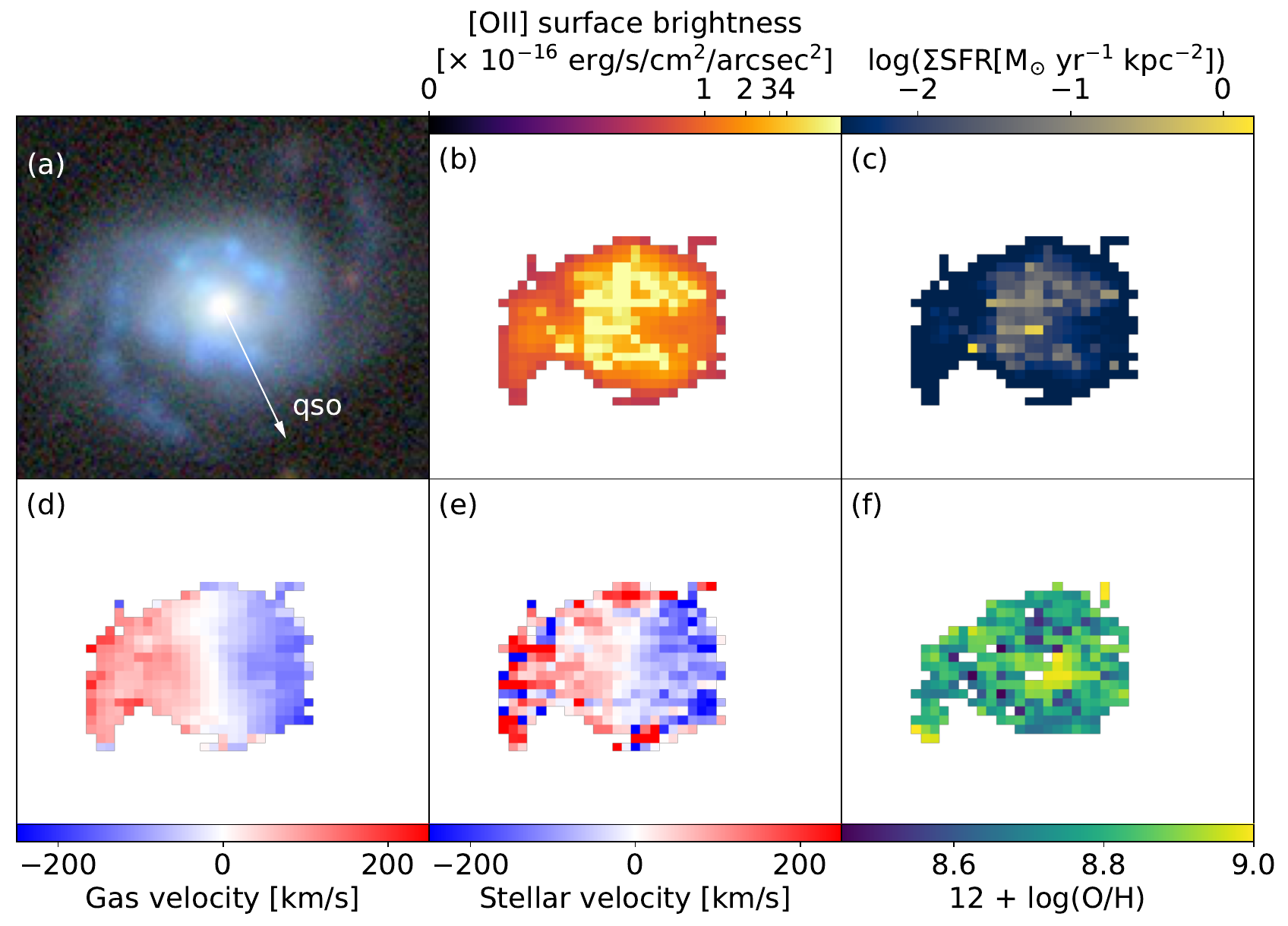}
    \caption{Maps of different properties of G1a: (a) DECaLS \textit{grz} image with the same orientations as Fig.~\ref{fig:system}. The white arrow points towards the location of the quasar, (b) {\OII} surface brightness, (c) $\Sigma_{\rm SFR}$, (d) gas velocity, (e) stellar velocity, and (f) gas metallicity.}
    \label{fig:face_on_maps}
\end{figure*}

\begin{figure*}
    \centering
    \includegraphics[width=\textwidth]{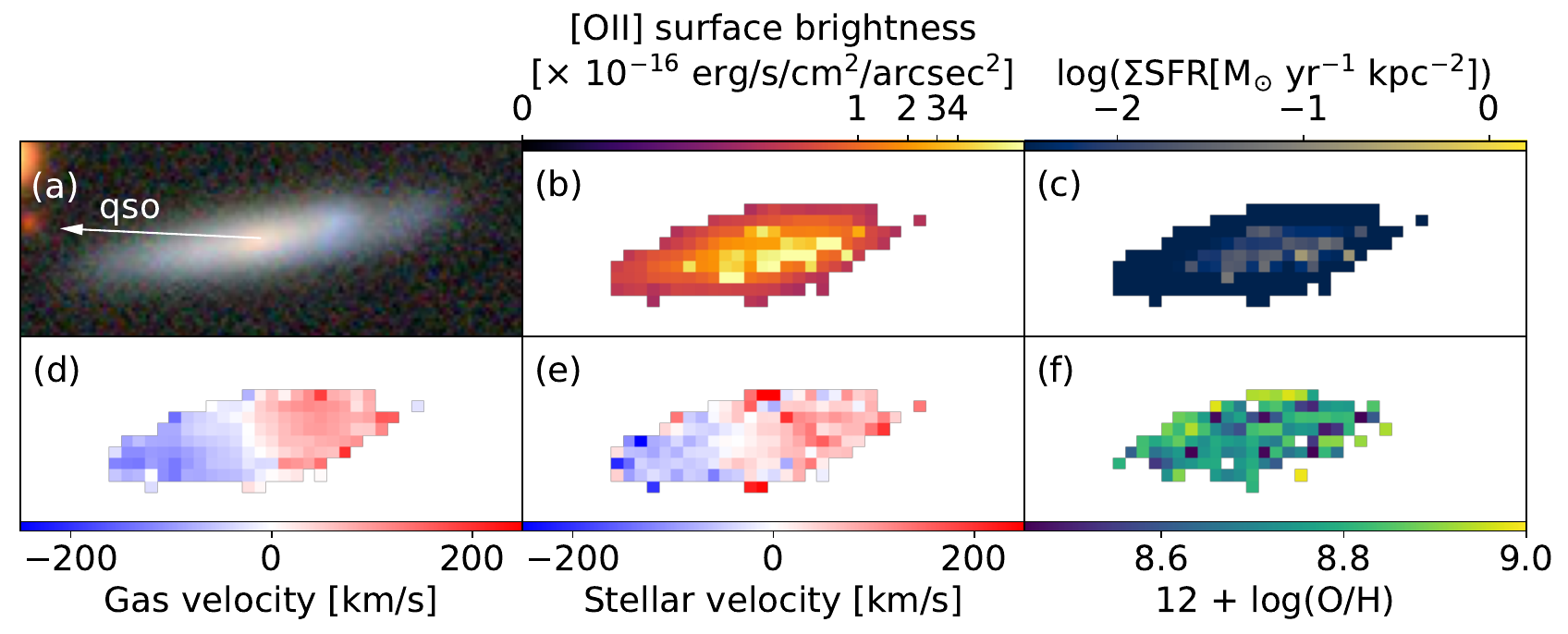}
    \caption{Maps of different properties of G1b: (a) DECaLS \textit{grz} image with the same orientation as Fig.~\ref{fig:system}. The white arrow points towards the location of the quasar, (b) {\OII} surface brightness, (c) $\Sigma_{\rm SFR}$, (d) gas velocity, (e) stellar velocity, and (f) gas metallicity.}
    \label{fig:edge_on_maps}
\end{figure*}

\begin{figure}

\begin{subfigure}[b]{\columnwidth}
    \centering
    \includegraphics[width=\columnwidth]{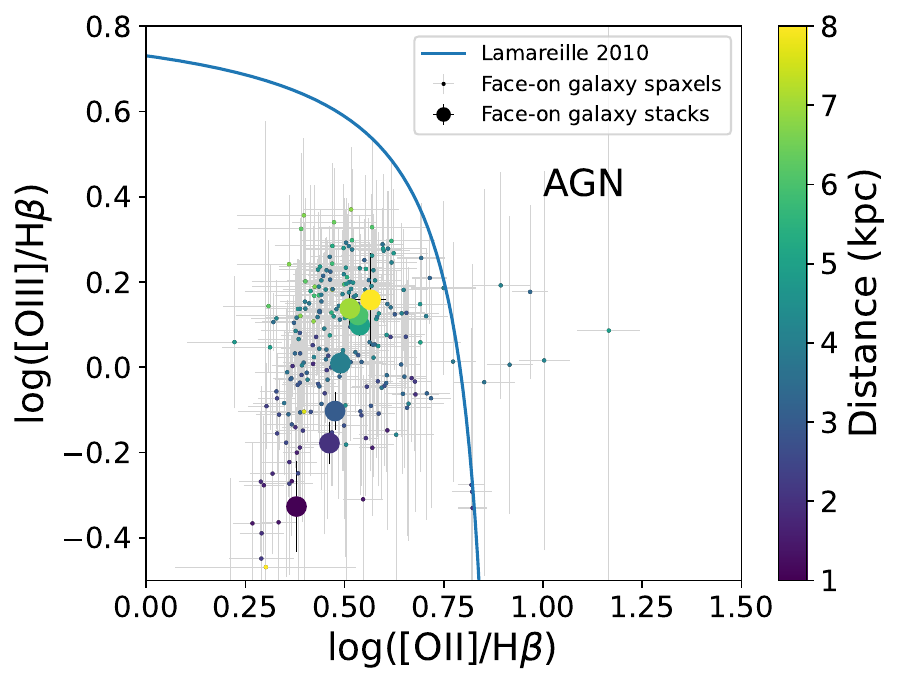}
\end{subfigure}

\begin{subfigure}[b]{\columnwidth}
    \centering
    \includegraphics[width=\columnwidth]{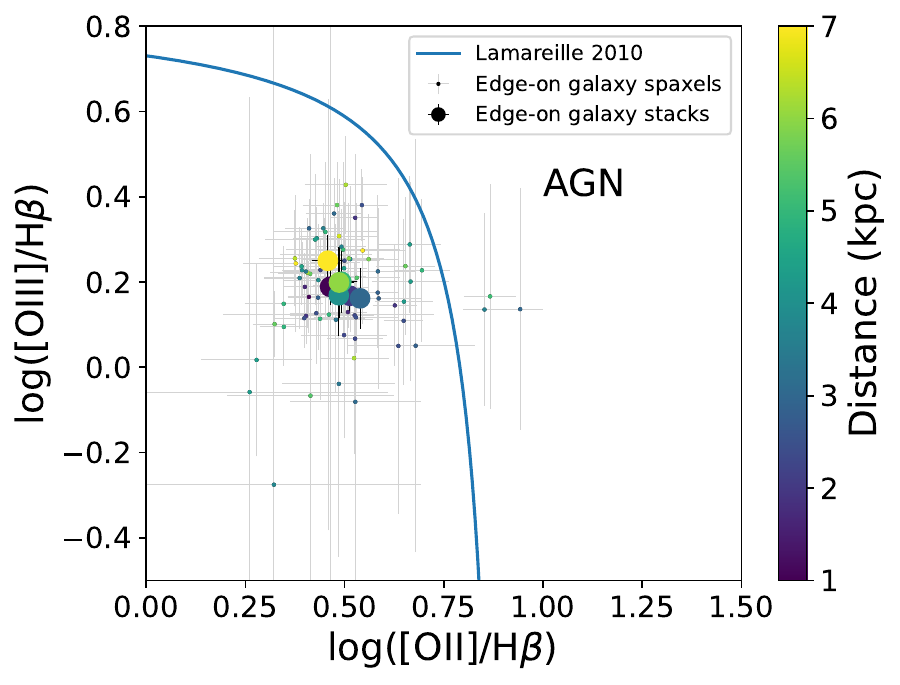}
\end{subfigure}

\caption{Line-ratio diagram of 1~kpc stacks in G1a (top) and G1b (bottom). The blue line separates the AGN and star-forming regions \citet{Lamareille2010}. Each stack is colour-coded by its distance to the centre of the galaxy. All stacks lie in the star-forming region.  The location of individual spaxels is displayed with smaller circles.}
\label{fig:bpt}
\end{figure}

\begin{figure*}
   \includegraphics[width=\textwidth]{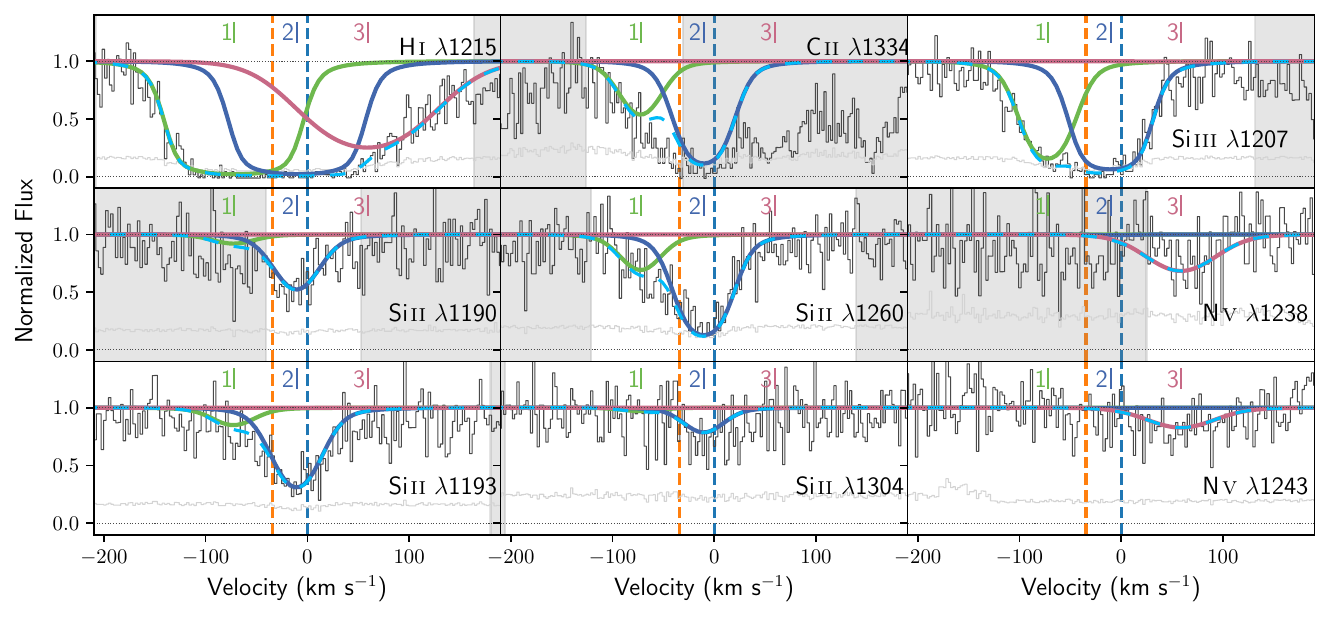}
   \caption{Absorption at the redshift ($z_{\rm abs}=0.04318$) of the distant pair, G1a and G1b, where strong H{\sc i}, Si{\sc ii}, Si{\sc iii} and N{\sc v} absorption is detected. The velocity zero-point is defined as the redshift of G1a and is represented by the vertical blue dashed line. Additionally, the redshift of G1b is indicated with the orange dashed line. Black and grey curves indicate the data and the $1\sigma$ uncertainty, respectively. Three clouds were fitted to this absorption system, where their profiles are colour-coded by thick solid curves and their velocity centroids are noted by the vertical ticks. The dashed cyan line indicates the total fit to the absorption profile. The grey-shaded areas were not considered in the modelling.}
   \label{fig:qso_abs_G1}
\end{figure*}

\begin{figure}
    \includegraphics[width=\columnwidth]{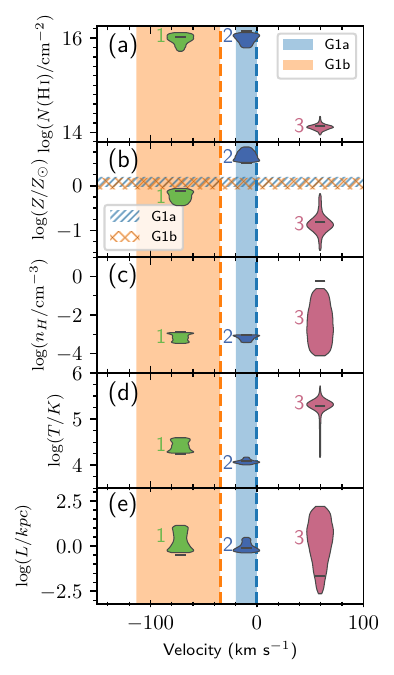}
    \caption{Properties of the absorption associated with the distant pair, G1a and G1b. The violin plots show the posterior distributions of (from top to bottom): column density, metallicity, hydrogen number density, temperature, and cloud thickness as a function of velocity. Each violin represents one of the velocity components of the absorption. The range of the violins indicates the 1$\sigma$ range for each parameter and the black horizontal lines represent their most likely value. The velocity zero point is defined by the redshift of G1a. The range of velocity values an extended rotating disk would have at the location of the quasar is represented by the blue-shaded area in the case of G1a, and the orange-shaded area in the case of G1b. The vertical dashed lines indicate the systemic velocity of each galaxy. The blue and orange hatched areas on panel (b) represent the metallicity range of G1a and G1b, respectively.}
    \label{fig:non_interacting_violins}
\end{figure}

\begin{figure}

\begin{subfigure}[b]{\columnwidth}
    \centering
    \includegraphics[width=\columnwidth]{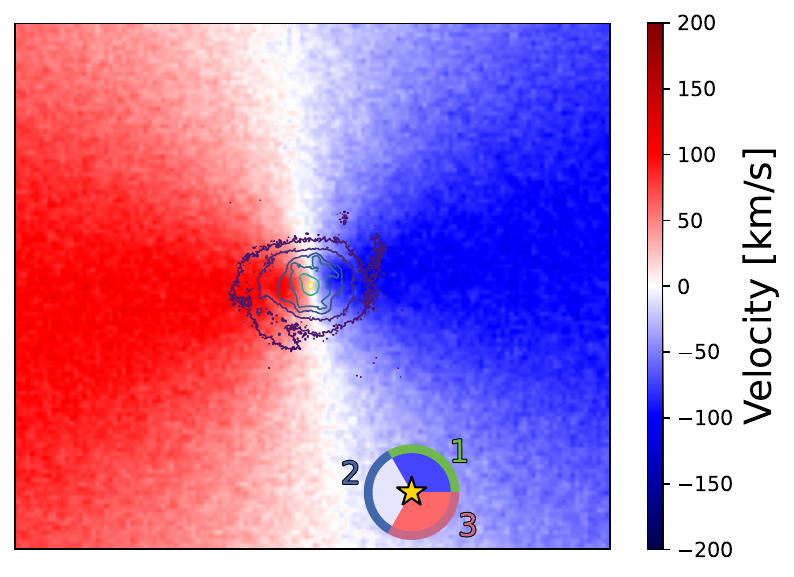}
\end{subfigure}

\begin{subfigure}[b]{\columnwidth}
    \centering
    \includegraphics[width=\columnwidth]{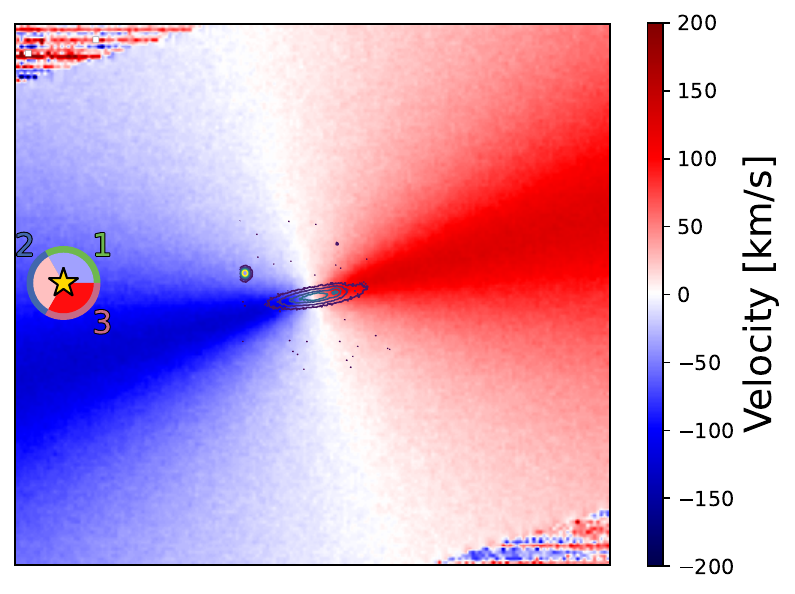}
\end{subfigure}

\caption{Extended rotation model of G1a (top) and G1b (bottom). The DECaLS g-band contours show the location of the galaxies. The yellow star represents the location of the quasar. The pie chart shows the velocities of the clouds found in the absorption, using the galaxies' redshifts as velocity zero-points. The outside part of the pie charts is colour-coded to match the colours used to represent the clouds in Figs.~\ref{fig:qso_abs_G1} and \ref{fig:non_interacting_violins}. In both cases, there is one cloud that matches the velocity range of the model at the location of the quasar (blue cloud for G1a and green cloud for G1b).}
\label{fig:models_distant}
\end{figure}

\begin{figure}
   \includegraphics[width=\columnwidth]{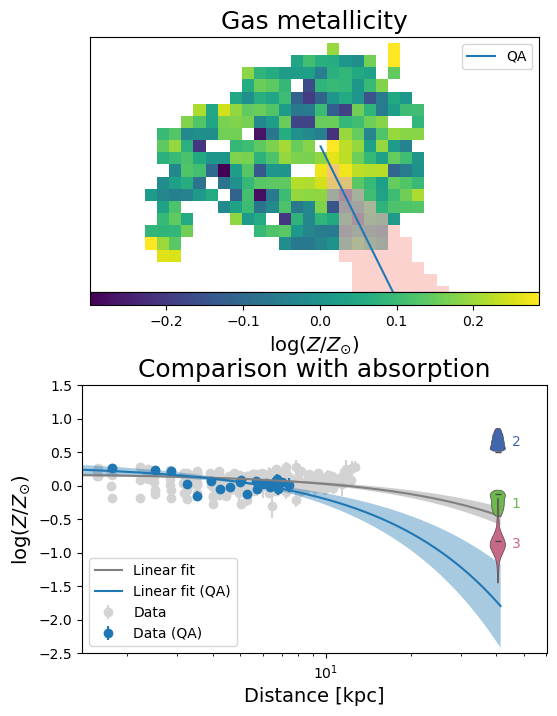}
   \caption{\textit{Left:} Gas metallicity map of G1a. The blue line connects the centre of G1a with the quasar and represents the quasar angle (QA). We selected the spaxels that were within $30^{\circ}$ of aperture from this line. The selected spaxels are highlighted in blue. \textit{Right:} Gas metallicity vs. distance. All of the spaxels are shown in grey, while those along the quasar direction aperture are shown in blue. We performed a linear fit to both of these samples, and the best fits are represented with the grey and blue lines, respectively. It is clear that the metallicity of G1a decreases with distance from its centre. The violin plots represent the metallicities of the different clouds in the CGM.}
   \label{fig:metallicity_gradient_G1}
\end{figure}

\begin{figure}
   \includegraphics[width=\columnwidth]{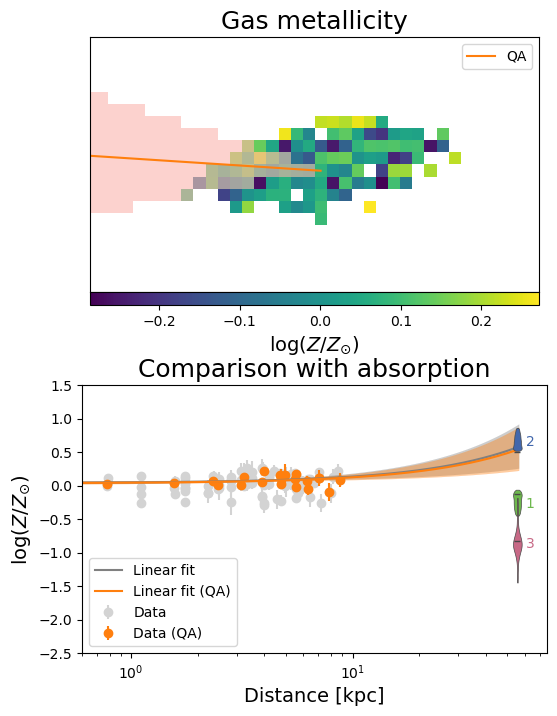}
   \caption{\textit{Left:} Gas metallicity map of G1b. The orange line connects the centre of G1b with the quasar and represents the quasar angle (QA). We selected the spaxels that were within $30^{\circ}$ of aperture from this line. The selected spaxels are highlighted in orange. \textit{Right:} Gas metallicity vs. distance. All of the spaxels are shown in grey, while those along the quasar direction aperture are shown in orange. We performed a linear fit to both of these samples, and the best fits are represented with the grey and orange lines, respectively. In this case, the slope of the fit is positive (0.010$\pm$0.006), but the large uncertainty indicates that the metallicity of G1b remains roughly constant with distance. This might be a consequence of its edge-on inclination. The violin plots represent the metallicities of the different clouds in the CGM.}
   \label{fig:metallicity_gradient_G2}
\end{figure}

\section{G1 -- Distant pair results} \label{distant_pair_results}

In Section~\ref{distant_pair_galaxy}, we present the galaxy properties (SFR, gas kinematics, stellar kinematics, metallicities, etc.) of G1a and G1b in the distant pair. We then present the CGM properties in Section~\ref{distant_pair_absorption}, followed by an analysis connecting the galaxy and CGM properties in Section~\ref{distant_pair_connection}. A further discussion of the origin of the CGM is presented in Section~\ref{distant_pair_origin}.

\subsection{G1 -- Galaxy properties} \label{distant_pair_galaxy}

G1 has two galaxies: one moderately inclined (G1a, see Fig.~\ref{fig:face_on_maps}(a), ${z_{\rm G1a}=0.04320}$) and another edge-on (G1b, see Fig.~\ref{fig:edge_on_maps}(a), ${z_{\rm G1b}=0.04308}$). Both have reasonably symmetric morphologies and rotation fields, which is an indication that they are not strongly interacting. General properties of these galaxies are listed in Table~\ref{tab:morphology}. They are at impact parameters of 40.4~kpc and 55.4~kpc from the quasar line-of-sight, respectively, while the two galaxies are separated from each other by 86.7~kpc (see Fig.~\ref{fig:system}). G1a is moderately inclined at $i=40^{\circ}$, while G1b is edge-on with an inclination angle of $i=89^{\circ}$. Both galaxies have strong {\OIIedblt} emission lines, as shown in Figs.~\ref{fig:face_on_maps}(b) and \ref{fig:edge_on_maps}(b). 

Rotation maps of the gas component are shown in Figs.~\ref{fig:face_on_maps}(d) and \ref{fig:edge_on_maps}(d), while those of the stellar component are shown in Figs.~\ref{fig:face_on_maps}(e) and \ref{fig:edge_on_maps}(d). The stars and gas in G1a are rotating in the same direction with similar rotation velocities, and this is also true for G1b. The coherence between the stars and gas in both galaxies, along with the symmetric and smooth rotation maps are further evidence that the two galaxies are not strongly interacting. G1a has a maximum gas rotation velocity of 237~km~s$^{-1}$ and G1b of 245~km~s$^{-1}$. Both galaxies have blueshifted rotation velocities towards the QSO line-of-sight.

Gas abundance maps can be found in Figs.~\ref{fig:face_on_maps}(f) and \ref{fig:edge_on_maps}(f). The median gas oxygen abundance of G1a is $12+\log({\rm O/H})=8.80\pm0.11$. Its abundance peaks close to the centre at $12+\log({\rm O/H})=9.00\pm0.06$ and decreases to $12+\log({\rm O/H})=8.41\pm0.18$ out to 6.5~kpc. The median oxygen abundance of G1b is $12+\log({\rm O/H})=8.76\pm0.13$, which is within uncertainties of G1a. In the case of G1b, we see low abundances along the major axis where accreting gas is expected to merge onto the disk, and high abundances along the minor axis where outflowing gas is ejected. These results are expected signatures of baryon cycle gas flows for star-forming galaxies. We explore the galaxy gradients and their relation to the CGM metallicities in Section~\ref{distant_pair_connection}.

None of the galaxies in G1 are particularly star-forming, with G1a having a SFR of 3.68~$\pm$~0.10~M$_{\astrosun}$~yr$^{-1}$ and G1b of 0.54~$\pm$~0.02~M$_{\astrosun}$~yr$^{-1}$. To explore the possibility of outflows occurring within the galaxies, we examine the $\Sigma_{\rm SFR}$. $\Sigma_{\rm SFR}$ are displayed in Figs.~\ref{fig:face_on_maps}(c) and \ref{fig:edge_on_maps}(c). The total $\Sigma_{\rm SFR}$ of G1a and G1b are 0.0245~$\substack{+0.02 \\ -0.01}$ and 0.0107~$\substack{+0.01 \\ -0.00}$~M$_{\astrosun}$~yr$^{-1}$~kpc$^{-2}$, respectively. In both cases, the $\Sigma_{\rm SFR}$ is higher near the centre and decreases with distance. Both G1a and G1b have a total $\Sigma_{\rm SFR}$ below the threshold required to produce strong star formation-driven outflows \citep[e.g.,][]{Heckman2015}. Nine spaxels in G1a have a $\Sigma_{\rm SFR}$ marginally higher than the threshold, with one spaxel reaching a maximum value of 1.41~$\pm$~0.15~M$_{\astrosun}$~yr$^{-1}$~kpc$^{-2}$. In the case of G1b, only two spaxels are above the threshold, going up to 0.27~$\pm$~0.02~M$_{\astrosun}$~yr$^{-1}$~kpc$^{-2}$. Therefore, it is possible that these small regions are currently producing outflows, but outflows are likely not occurring throughout the disks.

To determine whether these galaxies have AGN, we explore the line-ratio diagram \citep{Lamareille2010}. We produced line-ratio diagrams for each galaxy, which includes all spaxels with all lines detected above 5$\sigma$, as shown in Fig.~\ref{fig:bpt}. The figure shows that the vast majority of the galaxy is consistent with typical star-forming regions. Only six spaxels reside in the AGN region, however the 1$\sigma$ uncertainty bars on these points show the data is consistent with originating from star-forming regions. Furthermore, these points reside at distances of 5-6~kpc from the galaxy centre, so it is also unlikely they are connected to a central AGN. In the case of G1b, we see a similar trend as only three spaxels are in the AGN region, but they have large error bars and are still consistent with star-forming activity. Therefore, it is unlikely that these two galaxies currently have active AGN.

\subsection{G1 -- CGM absorption properties} \label{distant_pair_absorption}

We identified {\HI}, {\CII}, {\SiII}, {\SiIII}, and {\NV} absorption in the quasar spectrum. The rest-frame equivalent widths and column densities of these ions, as well as their individual cloud rest-frame equivalent widths and column densities, are detailed in Table~\ref{tab:abs_properties_G1}. The total {\HI} column density of the system is ${\log N({\HI}/{\rm cm}^{-2})=16.43 \substack{+0.03 \\ -0.04}}$. As shown in Fig.~\ref{fig:qso_abs_G1}, this system is determined to comprise three clouds, where two were constrained by the low/intermediate ionisation ions (clouds \textcolor{my_g1_green}{1} and \textcolor{my_g1_blue}{2} in Fig.~\ref{fig:qso_abs_G1}) and the third was constrained by the high ionisation ions (cloud \textcolor{my_pink}{3} in Fig.~\ref{fig:qso_abs_G1}). The absorption in {\HI} arises from both the low/intermediate and high ionisation phases, with most of the column density arising from the low/intermediate phase (${\log N({\HI}/{\rm cm}^{-2})=16.36 \substack{+0.05 \\ -0.05}}$).

\begin{table}
	\centering
	\caption{CGM absorption rest-frame equivalent widths and column densities for the distant pair.}
	\label{tab:abs_properties_G1}
	\begin{tabular}{lcccc} 
        \hline
        Ion & & v & EW & log(N / cm$^{-2}$) \\
         & & (km s$^{-1}$) & (\AA) &  \\
         (1)& (2) & (3) & (4) & (5) \\ 
        \hline
        {\HI} & Total &  & 1.09 $\pm$ 0.02 & $16.43_{-0.04}^{+0.03}$ \\[2pt]
         & \textcolor{my_g1_green}{1} & $-66.3_{-0.2}^{+0.4}$ &  0.59 $\pm$ 0.02 & $16.11_{-0.06}^{+0.05}$ \\[2pt]
         & \textcolor{my_g1_blue}{2} & $-3.8_{-0.6}^{+0.6}$ &  0.58 $\pm$ 0.02 & $16.15_{-0.05}^{+0.03}$ \\[2pt]
         & \textcolor{my_pink}{3} & $-65.4_{-2.6}^{+2.1}$ & 0.49 $\pm$ 0.02 & $14.13_{-0.03}^{+0.03}$ \\[2pt]
        {\CII} & Total & & 0.40 $\pm$ 0.03 & $14.60_{-0.02}^{+0.02}$ \\[2pt]
         & \textcolor{my_g1_green}{1} & $-66.3_{-0.2}^{+0.4}$ & 0.12 $\pm$ 0.03 & $13.90_{-0.03}^{+0.02}$ \\[2pt]
         & \textcolor{my_g1_blue}{2} & $-3.8_{-0.6}^{+0.6}$ &  0.29 $\pm$ 0.03 & $14.50_{-0.02}^{+0.02}$ \\[2pt]
         & \textcolor{my_pink}{3} & $-65.4_{-2.6}^{+2.1}$ & {\nodata} & $<12.96$ \\[2pt]
        {\SiII} & Total & & 0.76 $\pm$ 0.06 & $13.65_{-0.02}^{+0.02}$ \\[2pt]
         & \textcolor{my_g1_green}{1} & $-66.3_{-0.2}^{+0.4}$ & 0.14 $\pm$ 0.06  & $12.74_{-0.04}^{+0.05}$ \\[2pt]
         & \textcolor{my_g1_blue}{2} & $-3.8_{-0.6}^{+0.6}$ &  0.62 $\pm$ 0.06 & $13.59_{-0.02}^{+0.02}$ \\[2pt]
         & \textcolor{my_pink}{3} & $-65.4_{-2.6}^{+2.1}$ & {\nodata} & $<12.05$ \\[2pt]
        {\SiIII} & Total &  & 0.55 $\pm$ 0.02 & $14.11_{-0.04}^{+0.04}$ \\[2pt]
         & \textcolor{my_g1_green}{1} & $-66.3_{-0.2}^{+0.4}$ & 0.24 $\pm$ 0.02 & $13.45_{-0.02}^{+0.02}$ \\[2pt]
         & \textcolor{my_g1_blue}{2} & $-3.8_{-0.6}^{+0.6}$ &  0.36 $\pm$ 0.02 & $14.00_{-0.04}^{+0.05}$ \\[2pt]
         & \textcolor{my_pink}{3} & $-65.4_{-2.6}^{+2.1}$ & {\nodata} & $<11.88$ \\[2pt]
        {\NV} & Total &  & 0.18 $\pm$ 0.05 & $13.83_{-0.04}^{+0.04}$ \\[2pt]
         & \textcolor{my_g1_green}{1} & $-66.3_{-0.2}^{+0.4}$ & {\nodata} & $<13.10$ \\[2pt]
         & \textcolor{my_g1_blue}{2} & $-3.8_{-0.6}^{+0.6}$ & {\nodata} & $<13.10$ \\[2pt]
         & \textcolor{my_pink}{3} & $-65.4_{-2.6}^{+2.1}$ & 0.18 $\pm$ 0.05 & $13.83_{-0.04}^{+0.04}$ \\[2pt]
        \hline \\
	\end{tabular}

\raggedright The columns are: (1) ion, (2) component, (3) velocity of the cloud relative to $z = 0.04318$, (4) equivalent width, and (5) column density.
\end{table}

The velocities, hydrogen column densities, Doppler parameters, hydrogen densities, metallicities, temperatures, and sizes of each cloud are displayed in Fig.~\ref{fig:non_interacting_violins} and tabulated in Table~\ref{tab:vp_properties_G1}. The two clouds of the lower ionisation phase have column densities of ${\log N({\HI}/{\rm cm}^{-2})\approx16}$, metallicity comparable to or greater than solar, a size of $\leq 3$~kpc, hydrogen number densities of ${\log(n_{\rm H}/{\rm cm}^{-3})=-3}$, and temperatures of $T=10^{4}$~K. On the other hand, the higher ionisation component has a much lower column density at $\log N({\HI}/{\rm cm}^{-2})=14$, while at the same time, it has a much higher temperature of $T=10^{5.5}$~K. The temperature, size and hydrogen number density are unconstrained on this cloud.

\begin{table*}
\begin{center}
\begin{threeparttable}
  \setlength{\tabcolsep}{0.04in}
 \def\colhead#1{\multicolumn{1}{c}{#1}}
 \caption{Cloud-by-cloud properties of the absorption system associated with G1a and G1b \label{tab:vp_properties_G1}}

\begin{tabular}{lllllllllll}
    \hline\hline
      \colhead{(1)}     &
          \colhead{(2)}     &
          \colhead{(3)}     &
          \colhead{(4)}     &
          \colhead{(5)}     &
          \colhead{(6)}	    &
          \colhead{(7)}	    &
          \colhead{(8)}     &
           \colhead{(9)}   & 
           \colhead{(10)}   &
           \colhead{(11)} \\
            \colhead{Cloud}                &
          \colhead{$V$}           &
          \colhead{{\metallicity}}           &
          \colhead{{\hden}}           &
          \colhead{{\totalcolden}}           &          
          \colhead{{\colden}}           &
          \colhead{{\temp}}           &
          \colhead{{\thickness}} &
          \colhead{{\bturb}} &
          \colhead{{\btherm}} &
          \colhead{{\bnet}} \\
        \colhead{number}                &  
          \colhead{{\kms}}                &
          \colhead{}                &
          \colhead{}                &
          \colhead{}                &
        \colhead{}                &
          \colhead{} & 
        \colhead{} & 
          \colhead{{\kms}} & 
          \colhead{{\kms}} & 
          \colhead{{\kms}}  \\
    \hline 
          \textcolor{my_g1_green}{1} & $-66.3_{-0.2}^{+0.4}$ & $-0.16_{-0.05}^{+0.05}$ & $-2.83_{-0.04}^{+0.03}$ & $18.19_{-0.05}^{+0.05}$ & $16.11_{-0.06}^{+0.05}$ & $4.23_{-0.01}^{+0.02}$ & $-0.5_{-0.1}^{+0.1}$ & $22.3_{-0.2}^{+0.1}$ & $16.8_{-0.3}^{+0.3}$ & $27.9_{-0.2}^{+0.2}$\\[2pt]
         \textcolor{my_g1_blue}{2} & $-3.8_{-0.6}^{+0.6}$ & $0.45_{-0.02}^{+0.04}$ & $-2.95_{-0.03}^{+0.03}$ & $18.23_{-0.04}^{+0.04}$ & $16.15_{-0.05}^{+0.03}$ & $4.07_{-0.01}^{+0.01}$ & $-0.3_{-0.1}^{+0.1}$ & $25.0_{-0.5}^{+0.4}$ & $13.9_{-0.2}^{+0.2}$ & $28.6_{-0.4}^{+0.3}$\\[2pt]
         \textcolor{my_pink}{3} & $65.4_{-2.1}^{+2.6}$ & $<0.30$ & $>-4.00$ & $20.40_{-0.06}^{+0.06}$ & $14.13_{-0.03}^{+0.03}$ & $5.74_{-0.03}^{+0.03}$ & $<2.9$ & $31.3_{-9.0}^{+7.9}$ & $94.8_{-3.5}^{+3.5}$ & $100.1_{-3.8}^{+4.0}$\\  
        \hline
\end{tabular}
   
 Properties of the different components contributing to the absorption. Notes: (1) Cloud ID; (2) Velocity of the cloud; (3) metallicity of the cloud to the solar metallicity; (4) total hydrogen volume density of the cloud; (5) total hydrogen column density of the cloud; (6) neutral hydrogen column density of the cloud; (7) temperature of the cloud in kelvin; (8) inferred line of sight thickness of the cloud in kpc; (9) non-thermal Doppler broadening parameter of the cloud (10) thermal Doppler broadening parameter measured for {\HI}; (11) total Doppler broadening parameter measured for {\HI}. The marginalised posterior values of model parameters are given as the median along with the upper and lower bounds corresponding to the 16--84 percentiles.
 
\end{threeparttable}
\end{center}
 \end{table*}

\subsection{G1 -- Comparing the CGM and ISM} \label{distant_pair_connection}

We have presented the properties of the galaxies along with those of absorption in the previous two subsections. Here we explore the possible connections between the galaxies in the distant pair and their CGM.

First, we explore the kinematic connection between the galaxies and their CGM. Low ionisation absorption has been shown to be consistent with the co-rotation velocities of the galaxy \citep[e.g.][]{Steidel2010, Kacprzak_2010b, Ho2017, Rahmani_2018, Weng2023, Nateghi2023a, Nateghi2023b}. These models are displayed in Fig.~\ref{fig:models_distant}, with the contours showing the location of the galaxies, the yellow star displaying the position of the quasar and the pie charts presenting the velocities of the clouds found in the absorption (see Fig.~\ref{fig:qso_abs_G1} and Table~\ref{tab:vp_properties_G1}), using the galaxies' redshifts as the velocity zero-point. The colour of the outer ring of the pie chart corresponds to the colour of the specific clouds in all other figures. If we assume that the CGM gas is co-rotating, lagging or accreting into the ISM, we would expect to see the velocities of the absorption to lie between the galaxy systemic velocity and the velocity of the model in the location of the quasar \citep[e.g.,][]{Steidel2002}. The velocity ranges at the location of the quasar are shown in Fig.~\ref{fig:non_interacting_violins} as the blue-shaded area for G1a and the orange-shaded area for G1b. It is clear that two of the clouds match the velocity range of G1a and G1b (clouds \textcolor{my_g1_blue}{2} and \textcolor{my_g1_green}{1}, respectively). Both of these clouds are in the low-ionisation phase and have kinematics consistent with an accretion model. The higher ionisation component (cloud \textcolor{my_pink}{3}) is at higher velocities in the direction opposite to the rotation direction of both galaxies.

The range of metallicities of the ISM, defined as the median $\pm$ standard deviation, is shown in Fig.~\ref{fig:non_interacting_violins} as the blue hatched area for G1a and the orange hatched area for G1b. Metallicity gradients can be seen in Figs.~\ref{fig:metallicity_gradient_G1} and \ref{fig:metallicity_gradient_G2}. In the case of G1a, the metallicity decreases with distance, both in the direction of the quasar (with a slope of $-0.051\pm0.017$~dex~kpc$^{-1}$) and in general (with a slope of $-0.015\pm0.004$~dex~kpc$^{-1}$). On the other hand, the metallicity of G1b increases with distance in both the direction of the quasar and in general, but the slope is shallow ($0.010\pm0.006$~dex~kpc$^{-1}$), although this could be an effect of the galaxy's edge-on inclination, which makes the metallicity more likely to be a reflection of the outer regions of the disk at all radii. In the case of G1a, its average ISM metallicity is not consistent with any of the clouds, although it is within 1.5$\sigma$ of cloud \textcolor{my_g1_green}{1}. If we look at the gradients, cloud \textcolor{my_g1_green}{1} is consistent with the overall gradient, while cloud \textcolor{my_pink}{3} is consistent with the angular window gradient. Whereas in the case of G1b, none of the clouds are consistent with the average metallicity of the ISM, but the positive gradient makes it match with cloud \textcolor{my_g1_blue}{2} in both the direction of the quasar and as a general trend.

Altogether, cloud \textcolor{my_g1_blue}{2} is consistent with the rotation of G1a, while cloud \textcolor{my_g1_green}{1} is consistent with the rotation of G1b. However, cloud \textcolor{my_g1_blue}{2} is consistent with the metallicity gradient of G1b and cloud \textcolor{my_g1_green}{1} matches the metallicity gradient of G1a. The only high ionisation cloud, cloud \textcolor{my_pink}{3}, does not match the velocity or metallicity gradient of either galaxy in this pair.

\subsection{G1 -- Origin of the CGM absorption} \label{distant_pair_origin}

Given the properties of galaxies G1a and G1b, as well as the three clouds along the quasar sightline described in the previous sections, we attempt to ascertain the physical origin of these clouds.

First, we explore the possibilities of the absorption arising from outflows. The CGM originating from outflowing gas from G1b is unlikely given the star formation surface density of G1b, and the extremely large opening angle ($\theta/2>70$~degrees) required for the quasar to probe outflowing gas from G1b. G1a has a more optimal orientation for the quasar to probe its outflow. Following the methods of \citet{Ho&Martin2020} to deduce the sign of the disk inclination, we infer that we are viewing G1a such that a possible outflow in the direction of the quasar would be pointed away from the observer with redshifted velocities. Cloud \textcolor{my_pink}{3} is consistent with this scenario and has a velocity offset of 59~km~s$^{-1}$ with respect to G1a, which is within a reasonable range for outflow velocities \citep{Nielsen2017}. The properties of cloud \textcolor{my_pink}{3}  suggest that it is in higher ionisation phases, low column density, but potentially low metallicity. If cloud \textcolor{my_pink}{3} did originate from an outflow, then it no longer contains ISM metallicity and may have mixed with the ambient CGM.

We next explore whether some of the CGM absorption is consistent with being associated with co-rotating disks or gas accretion. For cloud \textcolor{my_g1_green}{1}, its velocity is consistent with gas losing angular momentum and being accreted onto G1b. Cloud \textcolor{my_g1_green}{1} also has a lower metallicity than the disk of the galaxy and has a lower metallicity than the extrapolated metallicity gradient. Thus, cloud \textcolor{my_g1_green}{1} is most consistent with originating from gas that is accreting onto G1b. Cloud \textcolor{my_pink}{3} is rotating in the opposite direction as G1a and G1b. However, this cloud aligns with the metallicity gradient of G1a. Cloud \textcolor{my_g1_blue}{2} has kinematics consistent with an accretion model with respect to G1a. However, the cloud metallicity is super-solar, which may imply that the gas was previously from an outflow that is being reaccreted into G1a. Given the face-on orientation of the galaxy, the quasar sightline likely probes a superposition of physical phenomena within the CGM.

We finally explore the possibility of gas arising from the intergalactic medium (IGM). Works such as \citet{Rudie2012}, \citet{Lehner2013}, \citet{Savage2014}, and \citet{Hafen2017} have shown that absorption of $\log N({\HI}/{\rm cm}^{-2})<14.5$ is likely associated with the IGM. While cloud \textcolor{my_pink}{3} is consistent with the extended metallicity gradient of G1a, it is not consistent with the rotation range of either galaxy. It is in a high ionisation state, has a high temperature, a low metallicity and a low column density. All of these properties are consistent with what we would observe from the IGM. This cloud might be associated with intergalactic gas along the line-of-sight of the quasar. Tidal streams are also a possibility for the origins of cloud~\textcolor{my_pink}{3}, but one would likely expect, given the direction of rotation of both galaxies, that velocities of that material would align with the rotation direction of the galaxies.

Overall, G1a and G1b seem like typical star-forming galaxies and they are consistent with their CGM having originated from a superposition of two galaxies with reasonable explanations for the potential origins of the individual clouds, such as recycled outflows (\textcolor{my_pink}{3}), accretion (\textcolor{my_g1_green}{1},\textcolor{my_g1_blue}{2}) or the IGM (\textcolor{my_pink}{3}). Therefore, they are a good control sample to compare with the interacting group which we discuss next.

\begin{figure}
    \centering
    \includegraphics[width=\columnwidth]{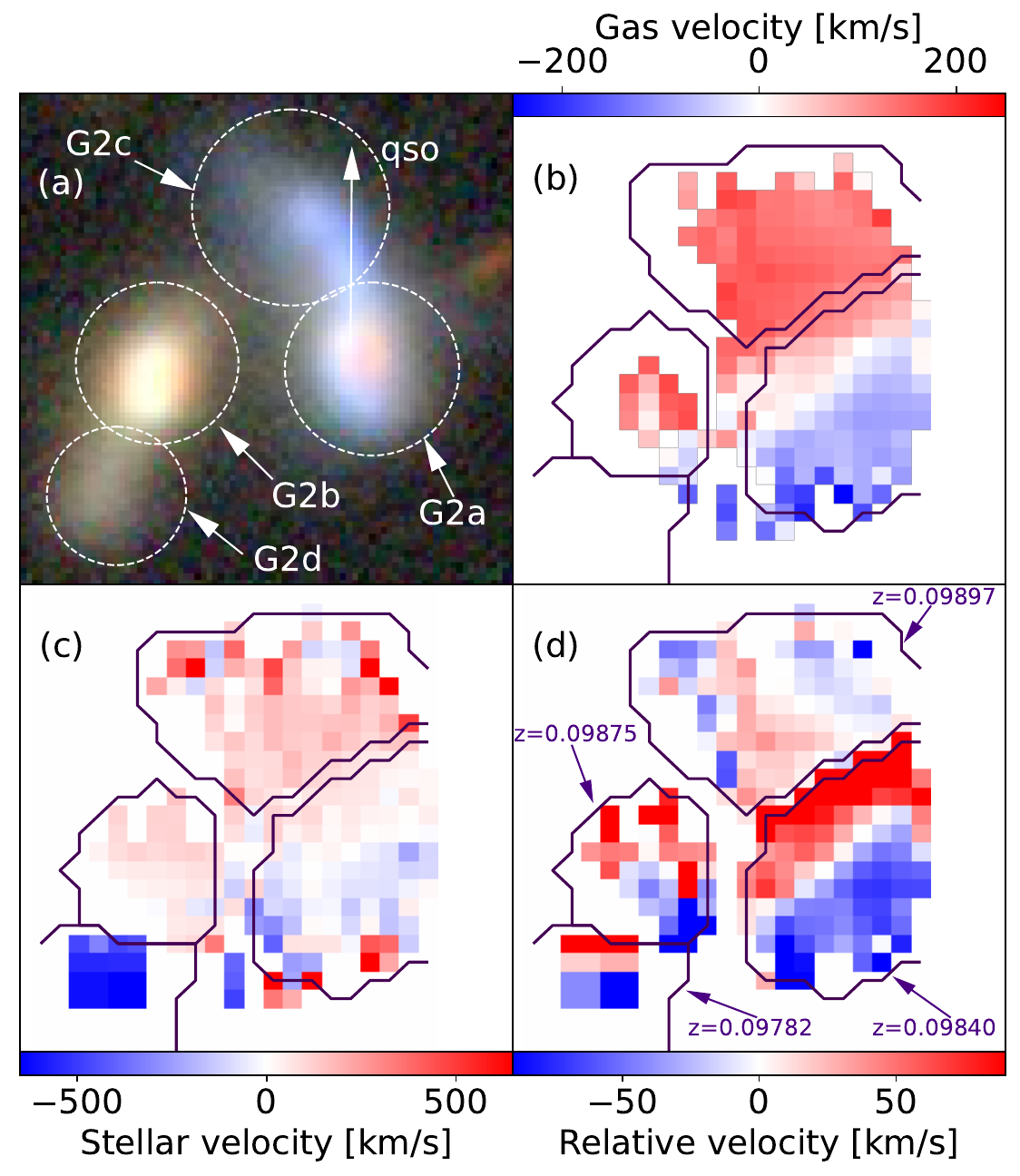}
    \caption{Maps of the merging system: (a) DECaLS \textit{grz} image with the same orientation as Fig.~\ref{fig:system}. The dashed circles indicate the different galaxies. (b) Gas velocity, using $z=0.09849$ as the velocity zero-point. (c) Stellar velocities, using $z=0.09849$ as the velocity zero-point. (d) Velocity of the galaxies using each of the galaxies' own redshift as the velocity zero-point. The velocity maps correspond to the gas velocity except in the case of G2d, which shows the stellar velocity. The purple contours indicate the spaxels associated with each galaxy.}
    \label{fig:G2_kin}
\end{figure}

\begin{figure}
    \centering
    \includegraphics[width=\columnwidth]{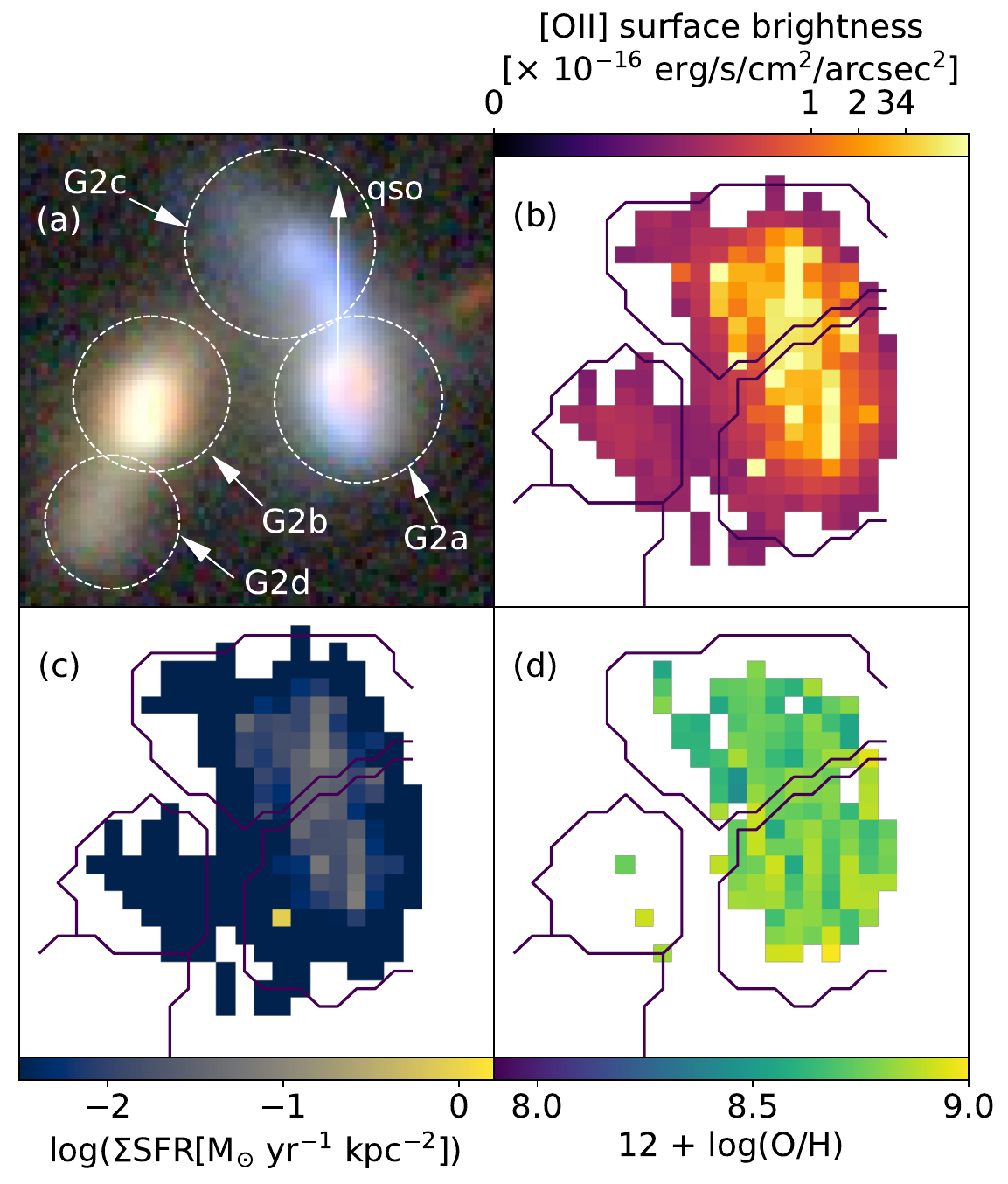}
    \caption{Maps of different properties of the merging system: (a) DECaLS \textit{grz} image with the same orientation as Fig.~\ref{fig:system}, (b) {\OIIe} surface brightness, (c) $\Sigma_{\rm SFR}$, and (d) gas metallicity. The purple contours indicate the spaxels associated with each galaxy.}
    \label{fig:G2_maps}
\end{figure}

\begin{figure}
   \includegraphics[width=\columnwidth]{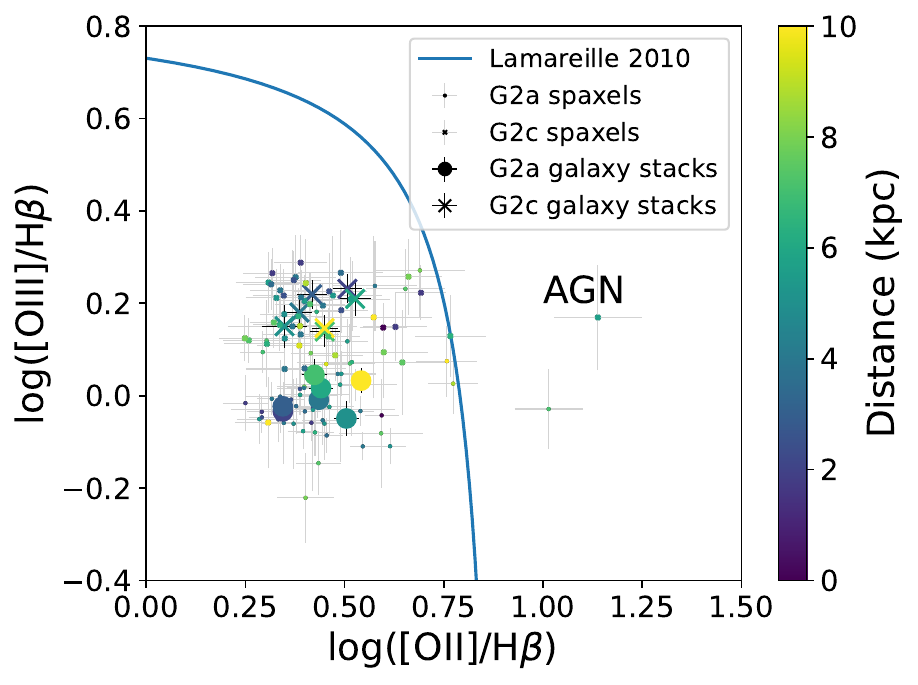}
   \caption{Line-ratio diagram of 1~kpc stacks in the merging system. Stacks belonging to G2a are represented with big circles, while the stacks belonging to G2c are represented with big crosses. Each stack is colour-coded by its distance to the centre of the galaxy. The separation curve between AGN and star-forming regions is the one derived in \citet{Lamareille2010}. Most stacks lie in the star-forming region. Only one spaxel is in the AGN region, but it is located in the bridge between G2a and G2c, therefore it is unlikely that it has AGN activity and it is more likely related to shocks produced by the merger. The location of individual spaxels is displayed with smaller circles and crosses.}
   \label{fig:bpt_merging}
\end{figure}

\begin{figure*}
   \includegraphics[width=\textwidth]{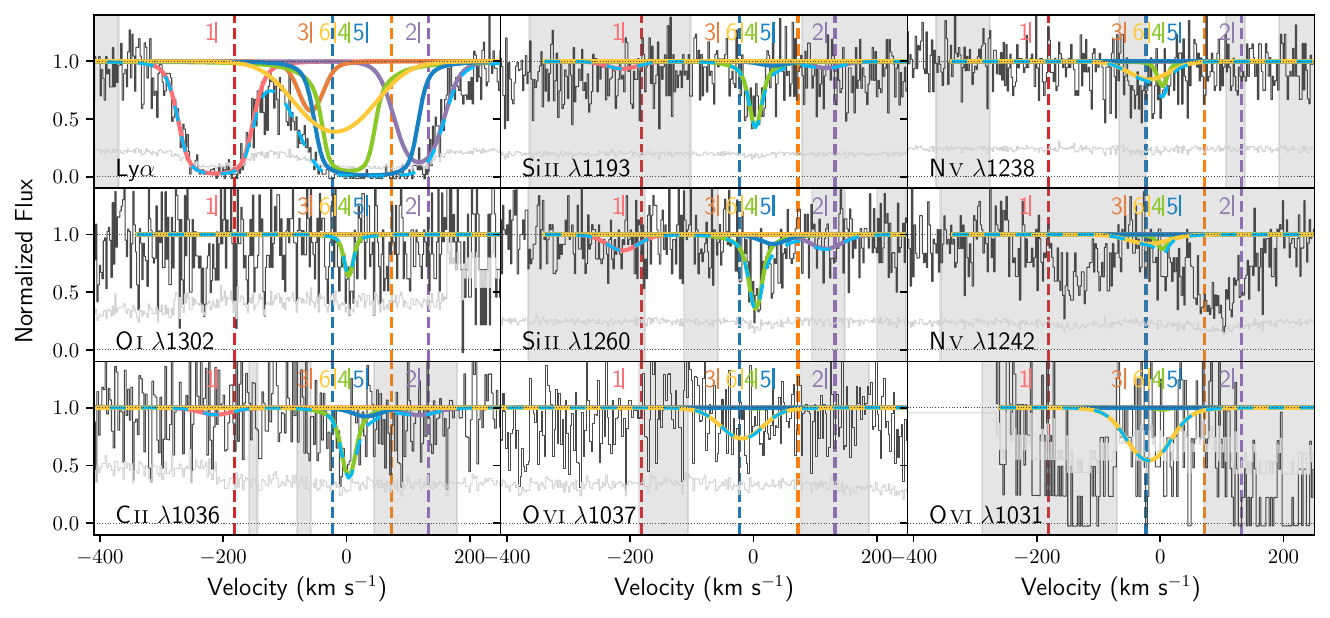}
   \caption{Absorption at the redshift ($z_{\rm abs}=0.098522$) of Group 2, where strong H{\sc i}, C{\sc ii}, Si{\sc ii}, Si{\sc iii}, N{\sc v} and O{\sc vi} absorption is detected, along with weaker C{\sc i}, N{\sc i} and O{\sc i} absorption. The velocity zero-point is defined as the redshift of the whole system, and the velocity of each galaxy is indicated with the vertical dashed lines. Black and grey curves indicate the data and error spectrum, respectively. Six clouds were fitted to this absorption system, where their profiles are colour-coded by thick solid curves and their velocity centroids are noted by the vertical ticks. The dashed cyan line indicates the total fit to the absorption profile. The grey-shaded areas were not considered in the modelling.}
   \label{fig:qso_abs_G2}
\end{figure*}

\begin{figure}
    \includegraphics[width=0.9\columnwidth]{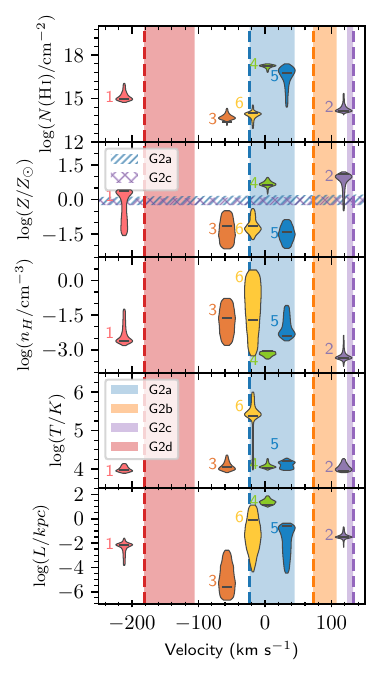}
    \caption{Properties of the CGM absorption associated with the merging system. The violin plots show the posterior distributions of cloud (from top to bottom): column density, metallicity, hydrogen number density, temperature, and thickness as a function of velocity. The velocity zero point is defined by the redshift of the whole system. Each violin represents one of the velocity components of the absorption centred at the redshift of the cloud. The vertical range of the darker and lighter violins indicates the $1\sigma$ and $3\sigma$ range for each parameter, respectively, and the black horizontal lines represent their most likely value. The width of each violin represents the posterior distribution of that particular absorption property. The vertical dashed lines indicate the systemic velocity of each galaxy. To compare the absorption to the galaxies, the range of velocity values an extended rotating galaxy disk would have at the location of the quasar is represented by the blue-shaded area in the case of G2a, the orange-shaded area for G2b, the purple-shaded area for G2c and the red-shaded area for G2d. The green shaded area represents the range of metallicities of the galaxies.}
    \label{fig:merging_violins}
\end{figure}

\begin{figure*}

      \begin{subfigure}[b]{0.47\textwidth}
          \centering
          \includegraphics[width=\textwidth]{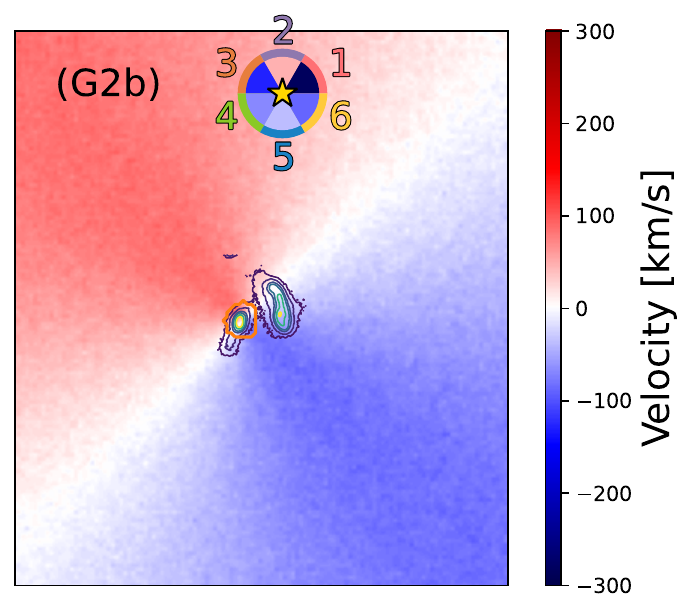}
      \end{subfigure}
      \hfill
      \begin{subfigure}[b]{0.47\textwidth}
          \centering
          \includegraphics[width=\textwidth]{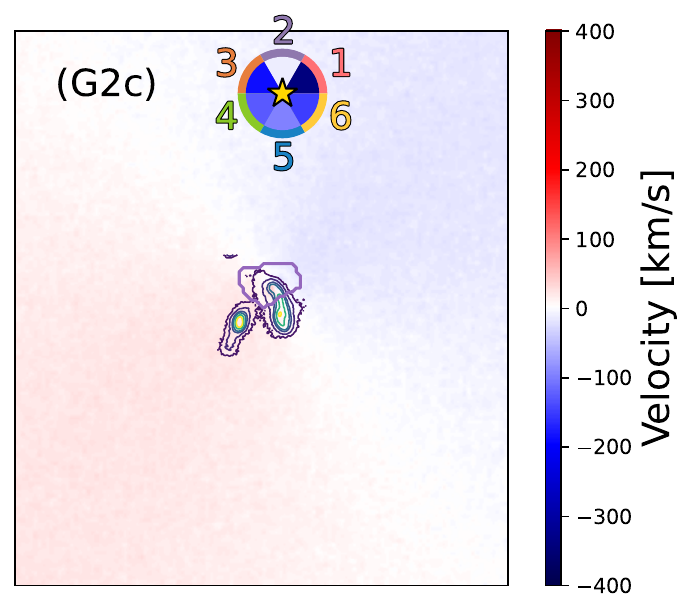}
      \end{subfigure}
      \begin{subfigure}[b]{0.47\textwidth}
          \centering
          \includegraphics[width=\textwidth]{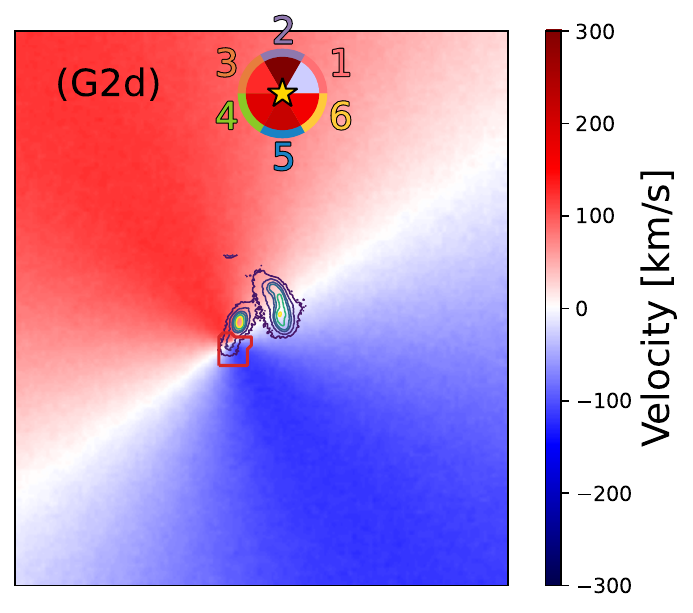}
      \end{subfigure}
      \hfill
      \begin{subfigure}[b]{0.47\textwidth}
          \centering
          \includegraphics[width=\textwidth]{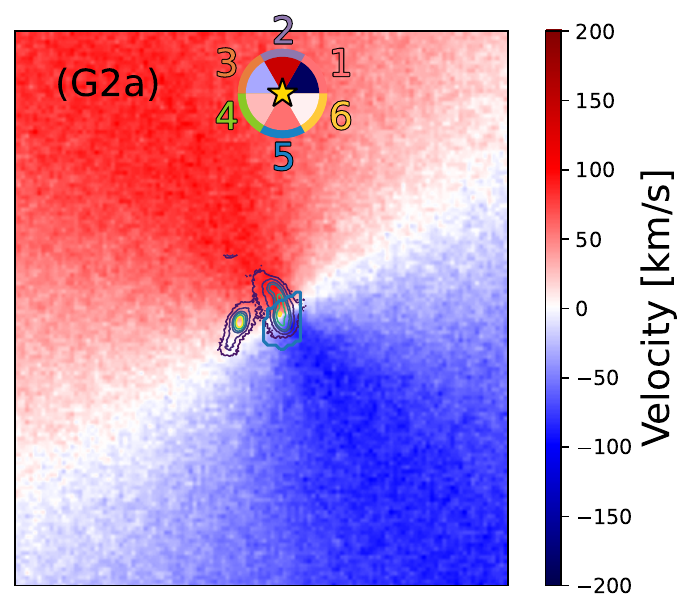}
      \end{subfigure}

\caption{Extended rotation models for the galaxies in the merging system. The DECaLS g-band contours indicate the location of the galaxies. The letter in the top left corner of the plots shows which galaxy the model belongs to, where the galaxy itself is also highlighted with yellow contours. The yellow star represents the location of the quasar. The pie chart shows the velocities of the clouds found in the absorption using the galaxies' redshifts as velocity zero-points. The outer annulus of the pie charts is colour-coded to match the colours used to represent the clouds in Figs.~\ref{fig:qso_abs_G2} and \ref{fig:merging_violins}.}
\label{fig:models_merging}
\end{figure*}

\begin{figure*}
   \includegraphics[width=\textwidth]{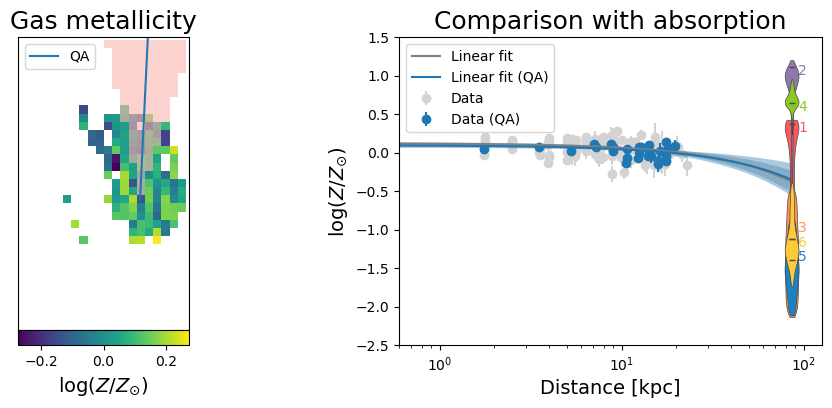}
   \caption{\textit{Left:} Gas metallicity map of the merging system. The blue line connects the centre of G2a with the quasar (labelled QA). To isolate the metallicity gradient in the direction of the quasar, we selected the spaxels that were within $30^{\circ}$ of aperture from this line and these spaxels are highlighted in blue. \textit{Right:} Gas metallicity vs. distance. All of the spaxels are shown in grey, while those along the quasar line-of-sight are shown in blue. We performed a linear fit to both of these samples and the best fits are represented with the grey and blue lines, respectively. It is clear that the metallicity of the merging system decreases with distance from its centre.  The violin plots represent the metallicities of the different clouds in the CGM.}
   \label{fig:metallicity_gradient_merging}
\end{figure*}

\section{G2 -- Merging system results} \label{merging_results}

In Section~\ref{merging_system_galaxy}, we present the galaxy properties (SFR, gas kinematics, stellar kinematics, among others) of G2a-d in the merging system. We then present the CGM properties in Section~\ref{merging_system_absorption}, followed by an analysis connecting the galaxy and CGM properties in Section~\ref{merging_system_both}. A further discussion of the origin of the CGM is presented in Section~\ref{merging_system_origin}.

\subsection{G2 -- Galaxy properties} \label{merging_system_galaxy}

This group (see Fig.~\ref{fig:system}) contains four merging galaxies that have a mean redshift of $z_{\rm G2}=0.09849$. While at first glance they may look like only two galaxies, further inspection of these system's redshifts (see Table~\ref{tab:morphology}) and kinematics (see Fig.~\ref{fig:G2_kin}) indicates that they are four distinct galaxies. This group has a velocity dispersion of 109$\pm$56~km~s$^{-1}$. Their distance to the QSO line-of-sight is 100~kpc.

G2a, G2b and G2c have strong {\OIIedblt} emission lines, as shown in Fig.~\ref{fig:G2_maps}(b). G2a and G2c dominate the emission from this group. For G2c, the {\OIIedblt} are the only emission lines detected. G2d does not have any emission lines. The only lines present in its spectrum are Ca~\textsc{ii}~H\&K absorption.

Kinematic maps of the ISM and stellar components of the entire group with the velocity zero-point set to $z_{\rm G2}$ are displayed in Figs.~\ref{fig:G2_kin}(b) and \ref{fig:G2_kin}(c). The whole system is rotating in the gas component. The stellar component appears to be rotating in the same direction as the gas as well and seems to have a similar range of velocities ($-$300 to 300~km~s$^{-1}$). It was not possible to calculate a gas velocity for G2d, as it does not feature any emission lines. In the case of the stellar kinematics, G2d appears quite blueshifted at $-$270~km~s$^{-1}$ compared to the rest of the group. Fig.~\ref{fig:G2_kin}(d) also shows the rotation kinematics of each galaxy using their own redshift as a velocity zero-point, allowing us to see the kinematics of each galaxy individually. Each of the galaxies is rotating in their own direction, while rotating around each other at the same time. When the kinematics of the merging system is modelled as a whole, the side closest to the quasar sightline is rotating with positive velocities. When modelling the kinematics of the individual galaxies, their maximum velocities are 152, 169, 93, and 254 ~km~s$^{-1}$ for G2a-d, respectively. While G2a,b,d appear to be rotating in the same direction as the group, G2c's rotation axis is perpendicular to the system and it has negative velocities towards the line-of-sight of the quasar. It is important to point out that the SNR of the spectra in G2d is considerably lower than that of the other galaxies, so performing Voronoi binning was necessary to reach a SNR~$\geq3$. Unfortunately, this meant we only had eight spaxels to analyse the kinematics of this galaxy, and the model had to extrapolate a maximum velocity with less information than the other galaxies. Thus, the model in this galaxy should be interpreted with caution. We further explore and model the kinematics of these galaxies in Section~\ref{merging_system_both}.

This group's ionised gas abundance map is shown in Fig.~\ref{fig:G2_maps}(d). \Hbeta~emission lines are quite weak, if present at all, on G2b, so we could only calculate its abundance in three spaxels. There are no \Hbeta~emission lines on G2d, so we could not calculate the abundance at all in that case. This is likely due to G2a and G2c stripping gas away from it during earlier stages of the merging. The median metallicity of G2a is $12+\log({\rm O/H})=8.79\pm0.09$, while that of G2c is $12+\log({\rm O/H})=8.73\pm0.10$, which is consistent with G2a within uncertainties. However, the gas metallicity is slightly higher where G2a and G2b meet, going up to $12+\log({\rm O/H})=8.93$ in that region.

We used pPXF to calculate the stellar ages of the galaxies. We found the median age of the stars in G2a to be $8.50\pm0.44$~Gyr, while the median age of the stars in G2c is $8.48\pm0.61$~Gyr. Therefore, the ages of the stars in G2a and G2c are consistent within uncertainties. However, in the case of G2b, the median age of its stars is $9.21\pm0.36$~Gyr, meaning that G2b's stellar ages are higher than G2a and G2c. Interestingly, the stars are younger where G2a and G2b meet, having ages as low as 7.48~Gyr in that area, suggesting a recent star-formation event likely triggered by the merging. Further evidence of a recent star-formation event is provided by the fact that the \Hdelta~emission line in G2b has a rest-frame equivalent width of 4.3~\AA, which is indicative of a post-starburst state \citep{Wu2018}. Unfortunately, G2d's spectra do not have enough SNR to perform a pPXF run successfully, so we do not have stellar ages for this galaxy. 

G2a has the highest SFR of all the galaxies in this Group at 5.13~$\pm$~0.70~M$_{\astrosun}$~yr$^{-1}$. G2c has a lower SFR at 2.30~$\pm$~0.03~M$_{\astrosun}$~yr$^{-1}$. In the case of G2b, we could only calculate its SFR$_{\OIIe}~=~$0.06~$\pm$~0.03~M$_{\astrosun}$~yr$^{-1}$, as it does not contain \Hbeta~emission lines. It was only possible for us to calculate an upper limit for the SFR of G2d as SFR~$\leq$~0.03~M$_{\astrosun}$~yr$^{-1}$. A $\Sigma_{\rm SFR}$ map of the group is shown in Fig.~\ref{fig:G2_maps}(c). The total $\Sigma_{\rm SFR}$ of G2a and G2c is 0.02~$\pm$0.01 and 0.01~$\pm0.01$~M$_{\astrosun}$~yr$^{-1}$~kpc$^{-2}$, respectively. For G2b, we calculated its {\OIIe} $\Sigma_{\rm SFR}$ as $0.06\pm0.03$~M$_{\astrosun}$~yr$^{-1}$~kpc$^{-2}$. The $\Sigma_{\rm SFR}$ is slightly higher where G2a and G2c meet, going as high as $0.9\pm0.2$~M$_{\astrosun}$~yr$^{-1}$~kpc$^{-2}$, although this spaxel is most likely an outlier. The $\Sigma_{\rm SFR}$ is higher than 0.1~M$_{\astrosun}$~yr$^{-1}$~kpc$^{-2}$ in only two spaxels in this system, so it is most likely not currently driving outflows as a whole, but might be driving them in those areas.

We produced line-ratio diagrams for each spaxel with all {\OIIe}, {\OIIIe} and \Hbeta~lines detected above 5$\sigma$, as shown in Fig.~\ref{fig:bpt_merging}. All of these spaxels are located in either G2a and G2c. The majority of the spaxels are consistent with star-forming regions. Three spaxels lie in the AGN region of the line-ratio diagram. However, not only are their uncertainties large enough for the points to still be consistent with the star-forming region, but they are also located in the bridge between G2a and G2c rather than at the galaxy centres. This area would also be more prone to shocks produced by the merger, making the line-ratios exhibit AGN-like characteristics, even without AGN activity \citep{DAgostino2019}. It is unlikely that an AGN would be located between the two galaxies, so we reject the hypothesis that either galaxy hosts an active AGN.

\subsection{G2 --  CGM absorption properties} \label{merging_system_absorption}

When examining the CGM absorption of the merging system, we identified H{\sc i}, N{\sc ii}, C{\sc ii}, Si{\sc ii}, N{\sc v} and O{\sc vi} absorption, along with modest O{\sc i} absorption in the quasar spectrum, as shown in Fig.~\ref{fig:qso_abs_G2}. The total {\HI} column density of the system is ${\log N({\HI}/{\rm cm}^{-2})=17.32 \substack{+0.09 \\ -0.11}}$. The line measurements for all of the detected ions are detailed in Table~\ref{tab:abs_properties_G2}. We modelled six components to this system, where five were constrained by low/intermediate ionisation ions (clouds \textcolor{my_red}{1}, \textcolor{my_purple}{2}, \textcolor{my_orange}{3}, \textcolor{my_green}{4} and \textcolor{my_blue}{5}) and a sixth was constrained by the high ionisation ions (cloud \textcolor{my_yellow}{6}). Two of the low ionisation components (clouds \textcolor{my_orange}{1} and \textcolor{my_blue}{2}) show no metals associated with them. In the case of cloud \textcolor{my_blue}{5}, the low metallicity is expected, given that it has a low column density. However, the column density of cloud \textcolor{my_orange}{3} is quite high, so it is interesting to see its low metallicity.

\begin{table}[h]
	\centering
	\caption{Equivalent widths and column densities for CGM absorption located at 100~kpc from the merging system.}
	\label{tab:abs_properties_G2}
	\begin{tabular}{lcccc}
        \hline
        Ion &  & v & EW & log(N / cm$^{-2}$) \\
         & & (km s$^{-1}$) & (\AA) & \\
         (1) & (2) & (3) & (4) & (5) \\
        \hline
        H {\sc i} & Total & & 1.67 $\pm$ 0.04 & $17.32_{-0.11}^{+0.09}$ \\[2pt]
         & \textcolor{my_red}{1} & $-221.9_{-0.7}^{+0.7}$ & 0.55 $\pm$ 0.04 & $14.99_{-0.16}^{+0.18}$ \\[2pt]
         & \textcolor{my_purple}{2} & $116.0_{-1.2}^{+1.4}$ & 0.36 $\pm$ 0.04 & $13.78_{-0.07}^{+0.07}$  \\[2pt]
         & \textcolor{my_orange}{3} & $-64.5_{-1.7}^{+1.4}$ & 0.10 $\pm$ 0.04 & $13.74_{-0.06}^{+0.06}$  \\[2pt]
         & \textcolor{my_green}{4} & $-6.1_{-0.5}^{+0.5}$  & 0.47 $\pm$ 0.04 & $17.23_{-0.09}^{+0.08}$ \\[2pt]
         & \textcolor{my_blue}{5} & $36.0_{-4.5}^{+5.0}$  & 0.77 $\pm$ 0.04 & $16.64_{-0.41}^{+0.35}$  \\[2pt]
         & \textcolor{my_yellow}{6} & $-25.8_{-5.2}^{+4.9}$  & 0.42 $\pm$ 0.04 & $13.81_{-0.10}^{+0.14}$ \\[2pt]
        C {\sc ii} & Total & & 0.13 $\pm$ 0.08 & $15.86_{-0.08}^{+0.08}$ \\[2pt]
         & \textcolor{my_red}{1} & $-221.9_{-0.7}^{+0.7}$ & 0.02 $\pm$ 0.08 & $<13.79$ \\[2pt]
         & \textcolor{my_purple}{2} & $116.0_{-1.2}^{+1.4}$ & 0.02 $\pm$ 0.08 & $<13.79$ \\[2pt]
         & \textcolor{my_orange}{3} & $-64.5_{-1.7}^{+1.4}$ & {\nodata} & $<13.79$ \\[2pt]
         & \textcolor{my_green}{4} & $-6.1_{-0.5}^{+0.5}$  & 0.09 $\pm$ 0.08 & $15.85_{-0.05}^{+0.04}$ \\[2pt]
         & \textcolor{my_blue}{5} & $36.0_{-4.5}^{+5.0}$  & 0.02 $\pm$ 0.08 & $<13.79$ \\[2pt]
         & \textcolor{my_yellow}{6} & $-25.8_{-5.2}^{+4.9}$ & {\nodata} & $<13.79$ \\[2pt]
        Si {\sc ii} & Total &  & 0.34 $\pm$ 0.05 & $14.90_{-0.06}^{+0.06}$ \\[2pt]
         & \textcolor{my_red}{1} & $-221.9_{-0.7}^{+0.7}$ & 0.06 $\pm$ 0.05 & $<12.19$ \\[2pt]
         & \textcolor{my_purple}{2} & $116.0_{-1.2}^{+1.4}$ & 0.15 $\pm$ 0.05 & $<12.19$ \\[2pt]
         & \textcolor{my_orange}{3} & $-64.5_{-1.7}^{+1.4}$ & {\nodata} & $<12.19$ \\[2pt]
         & \textcolor{my_green}{4} & $-6.1_{-0.5}^{+0.5}$  & 0.20 $\pm$ 0.05 & $14.90_{-0.06}^{+0.06}$ \\[2pt]
         & \textcolor{my_blue}{5} & $36.0_{-4.5}^{+5.0}$  & 0.03 $\pm$ 0.05 & $<12.19$ \\[2pt]
         & \textcolor{my_yellow}{6} & $-25.8_{-5.2}^{+4.9}$ & {\nodata} & $<12.19$ \\[2pt]
        Si {\sc iii} & Total &  & 0.39 $\pm$ 0.04 & $15.62_{-0.06}^{+0.05}$ \\[2pt]
         & \textcolor{my_red}{1} & $-221.9_{-0.7}^{+0.7}$ & 0.04 $\pm$ 0.04 & $<12.08$ \\[2pt]
         & \textcolor{my_purple}{2} & $116.0_{-1.2}^{+1.4}$ & 0.13 $\pm$ 0.04 & $12.77_{-0.04}^{+0.04}$ \\[2pt]
         & \textcolor{my_orange}{3} & $-64.5_{-1.7}^{+1.4}$ & {\nodata} & $<12.08$ \\[2pt]
         & \textcolor{my_green}{4} & $-6.1_{-0.5}^{+0.5}$  & 0.18 $\pm$ 0.04 & $15.62_{-0.06}^{+0.05}$ \\[2pt]
         & \textcolor{my_blue}{5} & $36.0_{-4.5}^{+5.0}$  & 0.06 $\pm$ 0.04 & $12.29_{-0.04}^{+0.04}$ \\[2pt]
         & \textcolor{my_yellow}{6} & $-25.8_{-5.2}^{+4.9}$ & {\nodata} & $<12.08$ \\[2pt]
        N {\sc v} & Total &  & 0.12 $\pm$ 0.06 & $13.61_{-0.05}^{+0.05}$ \\[2pt]
         & \textcolor{my_red}{1} & $-221.9_{-0.7}^{+0.7}$ & {\nodata} & $<13.26$ \\[2pt]
         & \textcolor{my_purple}{2} & $116.0_{-1.2}^{+1.4}$ & {\nodata} & $<13.26$ \\[2pt]
         & \textcolor{my_orange}{3} & $-64.5_{-1.7}^{+1.4}$ & {\nodata} & $<13.26$ \\[2pt]
         & \textcolor{my_green}{4} & $-6.1_{-0.5}^{+0.5}$  & 0.03 $\pm$ 0.06 & $13.54_{-0.10}^{+0.10}$ \\[2pt]
         & \textcolor{my_blue}{5} & $36.0_{-4.5}^{+5.0}$  & {\nodata} & $<13.26$ \\[2pt]
         & \textcolor{my_yellow}{6} & $-25.8_{-5.2}^{+4.9}$ & 0.09 $\pm$ 0.06 & \\[2pt]
        O {\sc vi} & Total &  & 0.25 $\pm$ 0.13 & $14.23_{-0.05}^{+0.05}$ \\[2pt]
         & \textcolor{my_red}{1} & $-221.9_{-0.7}^{+0.7}$ & {\nodata} & $<13.69$ \\[2pt]
         & \textcolor{my_purple}{2} & $116.0_{-1.2}^{+1.4}$ & {\nodata} & $<13.69$ \\[2pt]
         & \textcolor{my_orange}{3} & $-64.5_{-1.7}^{+1.4}$ & {\nodata} & $<13.69$ \\[2pt]
         & \textcolor{my_green}{4} & $-6.1_{-0.5}^{+0.5}$  & {\nodata} & $<13.69$ \\[2pt]
         & \textcolor{my_blue}{5} & $36.0_{-4.5}^{+5.0}$ & {\nodata} & $<13.69$ \\[2pt]
         & \textcolor{my_yellow}{6} & $-25.8_{-5.2}^{+4.9}$ & 0.25 $\pm$ 0.13 & $14.20_{-0.05}^{+0.05}$ \\
        \hline \\
 	\end{tabular}
  
  \raggedright The columns are: (1) ion, (2) component, (3) velocity of the cloud relative to $z = 0.098522$, (4) equivalent width, and (5) column density.
 \end{table}

The redshifts, hydrogen column densities, Doppler parameters, hydrogen densities, metallicities, temperatures and sizes of each cloud are displayed in Fig.~\ref{fig:merging_violins} and Table~\ref{tab:vp_properties_G2}. The three low-ionisation clouds with {\SiII} absorption (\textcolor{my_red}{1}, \textcolor{my_purple}{2} and \textcolor{my_green}{4}) have supersolar metallicities and a lower hydrogen number density of $\log (n_{\rm H}/{\rm cm}^{-3})\sim-3$, while the metallicities of the other two low-ionisation clouds (\textcolor{my_orange}{3} and \textcolor{my_blue}{5}) and the high-ionisation cloud (\textcolor{my_yellow}{6}) are subsolar and have a higher hydrogen number density ranging from $\log (n_{\rm H}/{\rm cm}^{-3})=-4$ to $-2.2$. All of the low ionisation clouds have lower temperatures of $10^{4}$~K, compared to the higher temperature of the higher ionisation cloud of $10^{5.5}$~K. The clouds have a wide range of sizes (ranging from $10^{-5}$ to 10~kpc) and column densities (ranging $13 \lesssim\log N({\HI}/{\rm cm}^{-2})\lesssim17$).

\begin{table*}
\begin{center}
\begin{threeparttable}
  \setlength{\tabcolsep}{0.04in}
 \def\colhead#1{\multicolumn{1}{c}{#1}}
 \caption{Cloud-by-cloud properties of the absorption system associated with merging pair \label{tab:vp_properties_G2}}

\begin{tabular}{lllllllllll}
    \hline\hline
      \colhead{(1)}     &
          \colhead{(2)}     &
          \colhead{(3)}     &
          \colhead{(4)}     &
          \colhead{(5)}     &
          \colhead{(6)}	    &
          \colhead{(7)}	    &
          \colhead{(8)}     &
           \colhead{(9)}   & 
           \colhead{(10)}   &
           \colhead{(11)} \\
            \colhead{Cloud}                &
          \colhead{$V$}           &
          \colhead{{\metallicity}}           &
          \colhead{{\hden}}           &
          \colhead{{\totalcolden}}           &          
          \colhead{{\colden}}           &
          \colhead{{\temp}}           &
          \colhead{{\thickness}} &
          \colhead{{\bturb}} &
          \colhead{{\btherm}} &
          \colhead{{\bnet}} \\
        \colhead{number}                &  
          \colhead{({\kms})}                &
          \colhead{}                &
          \colhead{}                &
          \colhead{}                &
        \colhead{}                &
          \colhead{} & 
        \colhead{} & 
          \colhead{({\kms})} & 
          \colhead{({\kms})} & 
          \colhead{({\kms})}  \\
    \hline 
          \textcolor{my_red}{1} & $-221.9_{-0.7}^{+0.7}$ & $<-0.18$ & $>-3.00$ & $15.81_{-0.49}^{+0.88}$ & $14.99_{-0.16}^{+0.18}$ & $4.04_{-0.06}^{+0.13}$ & $<-1.3$ & $28.1_{-2.6}^{+2.9}$ & $13.5_{-0.9}^{+2.1}$ & $31.4_{-2.3}^{+2.5}$\\[2pt]
          \textcolor{my_purple}{2} & $116.0_{-1.2}^{+1.4}$ & $<1.41$ & $>-4.04$ & $16.88_{-0.14}^{+0.15}$ & $13.78_{-0.07}^{+0.07}$ & $4.22_{-0.07}^{+0.07}$ & $<-0.4$ & $22.8_{-3.2}^{+2.9}$ & $16.5_{-1.3}^{+1.4}$ & $28.2_{-2.4}^{+2.3}$\\[2pt]
          \textcolor{my_orange}{3} & $-64.5_{-1.7}^{+1.4}$ & $<-0.05$ & $>-3.36$ & $14.61_{-0.55}^{+1.04}$ & $13.74_{-0.06}^{+0.06}$ & $4.04_{-0.08}^{+0.19}$ & $<-0.7$ & $19.3_{-5.0}^{+4.2}$ & $13.5_{-1.2}^{+3.2}$ & $24.2_{-3.5}^{+3.6}$\\[2pt]
          \textcolor{my_green}{4} & $-6.1_{-0.5}^{+0.5}$ & $0.64_{-0.07}^{+0.07}$ & $-3.18_{-0.03}^{+0.03}$ & $19.66_{-0.08}^{+0.07}$ & $17.23_{-0.09}^{+0.08}$ & $4.09_{-0.03}^{+0.03}$ & $1.3_{-0.1}^{+0.1}$ & $0.6_{-0.3}^{+0.3}$ & $14.2_{-0.4}^{+0.5}$ & $14.2_{-0.4}^{+0.5}$\\[2pt]
          \textcolor{my_blue}{5} & $36.0_{-4.5}^{+5.0}$ & $-0.98_{-0.35}^{+0.40}$ & $-2.15_{-0.12}^{+0.15}$ & $18.06_{-0.40}^{+0.34}$ & $16.64_{-0.41}^{+0.35}$ & $4.13_{-0.03}^{+0.03}$ & $-1.3_{-0.4}^{+0.4}$ & $28.9_{-2.7}^{+2.9}$ & $14.9_{-0.6}^{+0.5}$ & $32.5_{-2.5}^{+2.6}$\\[2pt]
          \textcolor{my_yellow}{6} & $-25.8_{-5.2}^{+4.9}$ & $<0.30$ & $>-4.00$ & $20.93_{-0.18}^{+0.14}$ & $13.81_{-0.14}^{+0.10}$ & $6.42_{-0.05}^{+0.03}$ & $<3.4$ & $32.6_{-12.4}^{+13.2}$ & $207.8_{-11.2}^{+8.5}$ & $211.0_{-10.8}^{+8.3}$\\  
          \hline
\end{tabular}
   
 Properties of the different components contributing to the absorption. Notes: (1) Cloud ID; (2) Velocity of the cloud; (3) metallicity of the cloud to the solar metallicity; (4) total hydrogen volume density of the cloud; (5) total hydrogen column density of the cloud; (6) neutral hydrogen column density of the cloud; (7) temperature of the cloud in kelvin; (8) inferred line of sight thickness of the cloud in kpc; (9) non-thermal Doppler broadening parameter of the cloud (10) thermal Doppler broadening parameter measured for {\HI}; (11) total Doppler broadening parameter measured for {\HI}. The marginalised posterior values of model parameters are given as the median along with the upper and lower bounds corresponding to the 16--84 percentiles.
 
\end{threeparttable}
\end{center}
 \end{table*}

\subsection{G2 -- Comparing CGM and ISM} \label{merging_system_both}

We have presented the properties of the galaxies along with those of absorption in the previous two subsections. Here we explore the possible connections between the galaxies in the merging system and their CGM.

The galaxy kinematic models are displayed in Fig.~\ref{fig:models_merging}. The contours in the figure highlight the location of the four galaxies and correspond to the DECaLS $g$-band image. Additionally, the yellow star shows the position of the quasar and the pie charts display the velocity of each of the clouds found in the absorption, with the outer part being colour-coded to match the colours of each cloud (see Fig.~\ref{fig:qso_abs_G2} and Table~\ref{tab:abs_properties_G2}), using the relative velocities of each galaxy. The velocity between the systemic velocity at the projected distance of the quasar is shown as shaded regions for each galaxy in Fig.~\ref{fig:merging_violins}. The vertical dashed lines represent the systemic velocity of each galaxy, colour-coded in the same way as the shaded areas. If the velocity of the absorbing gas is between the systemic velocity (vertical dashed lines) and the velocity of the model at the location of the quasar, that would be evidence that the absorbing gas is co-rotating, lagging or accreting onto the ISM \citep[e.g.,][]{Steidel2002}.  Good examples of this are G2a and G2c.  G2a has three clouds (\textcolor{my_green}{4}, \textcolor{my_blue}{5} and \textcolor{my_yellow}{6}) consistent with its rotation. Two of these clouds are in a low ionisation state but have higher column densities (clouds \textcolor{my_green}{4} and \textcolor{my_blue}{5}) while the other cloud (\textcolor{my_yellow}{6}) is in a high ionsation state with a low HI column density.  G2c has one cloud (\textcolor{my_purple}{2}) consistent with its rotation direction, but this cloud is at a higher velocity than the model at the quasar sightline. No clouds are consistent with corotation with G2d. However, it is important to note that the kinematics of these galaxies are unlikely to be described by rotation only. This means that although the models are useful for understanding the kinematics of the system, they might break down (e.g. if tidal streams or outflows are present).

The range of metallicities of the ISM (defined as the median of all the spaxels from Fig.~\ref{fig:G2_maps}, which is dominated by galaxies G2a and G2c) is shown in Fig.~\ref{fig:merging_violins} as the blue and purple hatched areas for G2a and G2c, respectively. To better understand the relationship of the CGM metallicity to the galaxy metallicities, we also modelled the metallicity gradients of galaxies G2a and G2c together in the direction of the quasar using a one-degree polynomial, as seen in Fig.~\ref{fig:metallicity_gradient_merging}. In this case, the metallicity decreases with distance, both in the direction of the quasar (with a slope of $-0.005\pm0.003$) and in general (with a slope of $-0.006\pm0.002$). This extended metallicity gradient is consistent with cloud \textcolor{my_orange}{3} of the absorption. Three components (clouds \textcolor{my_red}{1}, \textcolor{my_purple}{2} and \textcolor{my_green}{4}) have a metallicity that is consistent with the ISM within uncertainties. All these components are in a low-ionisation state. The other components have a metallicity that is lower than the ISM. Two of these components (clouds \textcolor{my_orange}{3} and \textcolor{my_blue}{5}) are in a low ionisation state, while the third one (cloud \textcolor{my_yellow}{6}) has a high ionisation state.

As a summary, we find that three clouds (\textcolor{my_green}{4}, \textcolor{my_blue}{5} and \textcolor{my_yellow}{6}) are consistent with co-rotation of G2a, and the other clouds are not consistent with co-rotating of any galaxy. Also, three clouds (\textcolor{my_red}{1}, \textcolor{my_purple}{2} and \textcolor{my_green}{4}) have a metallicity higher than, or equal to, the ISM, while the others have lower metallicities and are consistent with the metallicity gradient of G2a and G2c.

\subsection{G2 -- Origin of the CGM absorption} \label{merging_system_origin}

Given the properties of galaxies, as well as those of the clouds along the quasar sightline described in the previous sections, we attempt to ascertain the physical origin of each cloud. However, unlike G1 the merging process makes it more difficult to assign a particular galaxy to each of the clouds. This means that the superposition model does not match well with the observations of this group.

We do find clouds (\textcolor{my_green}{4}, \textcolor{my_blue}{5} and \textcolor{my_yellow}{6}) that have kinematics consistent with the rotation kinematics of G2a at metallicities at or below the ISM metallicity. However, it is highly unlikely that G2a would have a smooth extended rotating disk, which would have been disrupted during the merging/interacting process. 

Tidal streams are common in interacting galaxies, therefore we explore the possibility that our clouds are related to this process. Clouds \textcolor{my_red}{1} and \textcolor{my_orange}{3} do not match the rotation of any of the galaxies, although cloud \textcolor{my_orange}{3} is rotating in the same direction as G2d. It is also worth noting that clouds \textcolor{my_purple}{2}, \textcolor{my_green}{4} and \textcolor{my_blue}{5} are moving in the same direction as the merger. This is a suggestion that these clouds could not be corotating gas, and they might be tidal streams that are moving in the same direction as the merger.

Several studies have pointed out that the IGrM is clumpy and anisotropic \citep{Fossati2019, McCabe_2021b}. In the case of our clouds, we see that clouds \textcolor{my_red}{1}, \textcolor{my_orange}{3}, \textcolor{my_blue}{5} and \textcolor{my_yellow}{6} are consistent with a thickness of the order of 1~kpc and smaller. The small size of these clouds points to a clumpy CGM/IGrM scenario.

\section{Discussion} \label{discussion}

Both targets in the distant pair are consistent with being typical star-forming galaxies. The CGM absorption at $z = 0.04318$ has three velocity components The CGM absorption at $z = 0.098522$ is found to be associated with a merging system of galaxies. Three of the four galaxies in the merging system are also star-forming, but they have more complex kinematics. The absorption has six velocity components. In the next subsections we compare the properties and the CGM of the two groups. We also compare our results to those found in the literature.  

\subsection{Comparison of the two groups}

The galaxies in these two groups are in different stages of interaction. This translates into the galaxies and their CGM having different properties.

First, we compare the emission properties of the galaxies, starting from the SFR. Neither of the two groups seems to be particularly star-forming, although the spectrum of G2b is consistent with a post-starburst galaxy, given that its {\Hdelta} EW is 4.3~{\AA} \citep{Wu2018}. This is expected, given that there is evidence that mergers trigger star-burst episodes. Further proof that there was a recent star formation episode triggered by the merging is the fact that the stars in G2a are younger (8.5~Gyr) than the rest of the stars in the system (9.2~Gyr).

Next, we compare the variation between the absorption of the two systems. One major difference is that the merging system has six clouds, as opposed to the distant pair, which only has three, even though the impact parameter of the distant pair is much lower by a factor of two (G1: $D\simeq48$kpc and G2: $D\simeq100$kpc). This higher amount of components translates into having a higher \HI~equivalent width and column densities (G1: $\log N({\HI}/{\rm cm}^{-2})=16.43$ and G2: ${\log N({\HI}/{\rm cm}^{-2})=17.33}$). This is inconsistent with the anti-correlation between {\HI} and $D$ found for isolated galaxies \citep[][and references therein]{Kacprzak2021}. However, this is consistent with previous studies that show that the CGM of group galaxies extends into further distances \citep{Bordoloi_2011a, Nielsen_2018b}. This increase in absorption strength appears to increase even more in mergers. Additionally, G2 has a larger number of clouds at sub-kpc scales. Moreover, the smaller size of one of the clouds in G1 is most likely associated with the IGM, rather than the CGM of the whole system. This difference in the sizes of the clouds in both systems could be due to the merging creating smaller clumps or small scale structure in the CGM of G2 \citep{Hani_2018, Sparre2022}.

Looking at the kinematics of both groups, we note that the merging system has a wider range of velocities. The absorption on the distant pair only spans a range of $\sim$200~km~s$^{-1}$, compared to the $\sim$400~km~s$^{-1}$ of the merging system. This is consistent with \citet{Nielsen_2018b}, who showed that the CGM of group galaxies is more kinematically complex. Our results suggest that this level of kinematic complexity likely increases even more for merging galaxies. Another interesting difference between the two groups is that the gas in the merging system is in a higher ionisation state. This is expected, as the merging would create the ideal conditions to ionise the gas further than usual and could be situated in a higher halo mass system, further populating the level of ionised gas \citep{Oppenheimer2016, Oppenheimer2021, Wijers2022}.  

We can also explore the differences in the connections between ISM and CGM of both groups. The ISM metallicities of the galaxies in G1 are comparable to those in G2, both in their ISM and CGM. It is interesting to note that we see a much wider range of metallicities in the merging system, spanning 1.62~dex, compared to those of the distant pair, which span only 0.61~dex. The lower metallicity clouds in the merging pair seem to match the rotation of the system, unlike the low metallicity cloud in the distant pair, which does not seem to match any of the rotation. This suggests that the gas has different origins and again points to a scenario that merging galaxies produce a more complex CGM. 

As a general conclusion, the CGM of G2 seems to be more complex than the CGM of G1. This is expected because G2 is in a more advanced state of merging, so the CGM should be more complicated.

\subsection{Comparison with the literature}

It has been shown that absorption of group galaxies tends to be stronger and extend to further distances \citep{Chen_2010a, Kacprzak_2010a, Nielsen_2018b, Hamanowicz_2020, Cherrey2023}. While clear differences in the properties of the CGM exist between group and isolated galaxies, very few studies have well-quantified galaxy group selection criteria. Thus, how exactly different stages of galaxy environment affect the CGM remains less clear. Furthermore, group galaxies have been proposed as an explanation for the existence of {\MgII} absorbers with equivalent width higher than 2~\AA, also known as ultrastrong absorbers \citep{Nestor2007, Guha2022}. Our study goes one step further by comparing the absorption of group galaxies that are not strongly interacting with those that are merging. We find both equivalent widths and column densities are consistently stronger in G2, even though the QSO probes a larger distance in this case, suggesting that the merging exacerbates the enhancement in absorption strengths and extents.

\citet{Bordoloi_2011a} presented the superposition model to explain the distribution of CGM in group galaxies. This model assumes that the absorption of the group can be explained as the superposition of the absorption of each of the individual galaxies of the group. In the case of G1, we can identify which absorption component most likely corresponds to which galaxy, supporting this model even further. However, this separation is not as clear in the case of G2, indicating that this model only works well in the case of non strongly interacting galaxies, or galaxies that are in the early stages of interaction. This effect could be due to the gas having more opportunities to mix in more strongly interacting or merging systems so that the gas then loses its kinematic and metallicity connection to the original host. Additionally, other papers have suggested different approaches to the superposition model. \citet{Nielsen_2018b} found their sample does not fit the superposition model, based on the kinematic spread of the absorption systems and suggested a shared CGM model instead. Alternatively, \citet{Fossati2019} attributed the stronger absorption to filaments and tidal streams, while \citet{Dutta2020} proposed a mix of the superposition model with interactions between the individual galaxies. \citet{Beckett2023} developed a model for the CGM of group galaxies that includes outflows and disks, but they were unable to apply it to a merging system. These results, along with other studies that find anisotropy in the CGM of group galaxies \citep{Butsky2019, McCabe_2021b}, call for the need to develop more complex models or detailed high-resolution simulations to explain the CGM distribution of interacting and merging galaxies.

\citet{Nielsen_2018b} compared the kinematics of isolated and group galaxies by calculating their pixel-velocity Two-Point Correlation Function (TPCF). This function is defined as the probability distribution of the separation in velocity of every pair of absorbing pixels in the sample. They found that the TPCF of group galaxies has more power at higher velocities, suggesting not only that their absorption is more kinematically complex, but also that this trend is originating through galaxy--galaxy interactions and/or major merger events. In the case of our data, we find that the merging system has wider absorption profiles (400~km~s$^{-1}$ in the merging system compared to 200~km~s$^{-1}$ in distant pair) and double the velocity components, compared to the distant pair. Our results further suggest that the reason the CGM of groups is more kinematically complex is because of the interactions between the group galaxies. Additionally, \citet{Nielsen_2018b} find that group galaxies tend to have a higher fraction of clouds with a velocity offset higher than 100~km~s$^{-1}$, also known as high-velocity clouds. In the case of our systems, the merging group has two of these high-velocity clouds, as opposed to the distant pair which has zero. This indicates that one of the potential causes for this trend is the CGM being more highly kinematically distributed resulting from the interaction of the galaxies in the group.

Multiple works have found a decreased \CIV~\citep{Burchett2016} and \OVI~\citep{Stocke2013, Pointon_2017a, Ng2019, McCabe_2021b} absorption associated with group galaxies, and have come to the conclusion that this is due to the higher temperature in group environments creating the perfect conditions to ionise the gas even further. Our study does not have the data to analyse the hot phase of the CGM around our galaxies, but we have found that the CGM of G2 is in a higher ionisation state, given the presence of \NV~and \OVI~in cloud \textcolor{my_yellow}{6}. Although \NV~is also present in the absorption associated with the distant pair, this absorption is most likely a part of the IGM, and not related to the CGM of the galaxies in this system. This result points to the idea that mergers could contribute to gas heating in group environments, and ionising the gas surrounding them.

In their simulation study, \citet{Hani_2018} found that the metallicity of the CGM of post-merger systems is higher than in pre-mergers. This is due to mergers being outflow-dominated and suggests metallicity in group environments should be higher. However, observational studies have mixed results on this subject; while some find that the absorption metallicity in group systems is higher than in isolated galaxies \citep{Lehner2017, Beckett2023}, others find non-significant differences \citep{Pointon_2020}. In the case of our galaxies, we find that the merging system spans a wider range of metallicities than the distant pair, spanning 1.62~dex, compared to the distant pair which only spans 0.61~dex. Although we do not see current active outflows, we still observe higher metallicities at higher impact parameters in the CGM of merging galaxies. This evidence, in addition to the post-starburst state of G2b, suggests that G2 was dominated by outflows at the early stages of merging. However, we find lower metallicity clouds in this system as well, suggesting that the full picture is even more complex.

Moreover, the metallicity gradients also tell a story about the origin of the gas. Although the metallicity of the absorption tends to be lower than that of the ISM \citep{Peroux2016, Kacprzak2019}, some studies have shown that galaxies with low metallicity in their centres tend to have higher metallicities in their CGM \citep{Kulkarni2019}. The authors of this later study attribute this trend to dilution produced by mergers and infalling gas. This picture is different in group environments, where \citet{Hamanowicz_2020} find no significant metallicity gradient between the ISM and the CGM of the groups in their sample, linking this result with the fact that their absorbers are associated with multiple galaxies. In the case of our galaxies, we have mixed results. On the one hand, G1a has a metallicity gradient with a negative slope, but G1b has a slightly positive one. On the other hand, the galaxies in G2 have negative slopes in their metallicity gradients. This could indicate that the galaxies in the distant pair are actually going through early stages of interaction that are not visible in emission.

\section{Summary and conclusions} \label{conclusions}

In this paper, we studied two galaxy systems: one with two galaxies that are not visibly interacting (G1), and a second one with four merging galaxies (G2). Both of these systems are along the line of sight of the same QSO. Using Keck/KCWI and HST/COS data, we calculated both the ISM and CGM properties of all the galaxies in these systems, and found the following:

\begin{itemize}
    \item G1 consists of two galaxies at $z=0.043$, G1a and G1b, which have impact parameters of 40.4 and 55.5~kpc, respectively. The galaxies of this system are rotating with maximum velocities of $\sim$250~km~s$^{-1}$ and have symmetrical velocity maps, confirming that these galaxies are not strongly interacting with each other. Both galaxies in this group have a median oxygen abundance of {${12+\log({\rm O/H})=8.8}$}. The total $\Sigma_{\rm SFR}$ of G1a and G1b are 0.02 and 0.01~M$_{\astrosun}$~yr$^{-1}$~kpc$^{-2}$, so it is unlikely that they are driving outflows. Additionally, there is no evidence of AGN activity in these galaxies.
    \item In the case of G2, it features four galaxies at $z=0.098$, G2a-d, and has an impact parameter of 97.7~kpc. The galaxies of G2 seem to be rotating in the same direction, but when we analysed the kinematics of each of them separately, we found they exhibit a much more complicated behaviour: the galaxies are rotating around each other while rotating on their own axes as well. The group also spans a wide range of velocities, having a velocity dispersion of 109~km~s$^{-1}$. G2a and G2c have median oxygen abundances of {$12+\log({\rm O/H})=8.8$} and 8.7, respectively. We could not calculate oxygen abundances in the other two galaxies, as they do not have any Hydrogen emission lines. There is evidence of a recent star-formation event in this system, given that the median age of the stars in the bridge between G2a and G2b is considerably lower (8.5~Gyr) than that of the rest of the group (9.2~Gyr). Additionally, the spectrum of G2b shows signs of being a post-starburst galaxy. The total $\Sigma_{\rm SFR}$ of G2a and G2c are 0.02 and 0.01~M$_{\astrosun}$~yr$^{-1}$~kpc$^{-2}$, respectively, so it is unlikely that they currently are driving outflows. Additionally, there is no evidence of AGN activity in this system.
    \item We have identified {\HI}, {\CII}, {\SiII}, {\SiIII}, and {\NV} absorption in the quasar spectrum at the redshift of G1, having a total $\log (N({\HI})/{\rm cm}^{-2})=16.43$. This absorption is divided into three velocity components, or clouds, spanning a velocity range of 132~km~s$^{-1}$. Two of these components are in a lower ionisation phase, while the other one is in a higher ionisation phase. The two clouds of the lower ionisation phase have column densities of ${\log (N({\HI})/{\rm cm}^{-2})\approx16}$, metallicity comparable to or greater than solar, a size of $\leq 3$~kpc, hydrogen number densities of $\log(n_{\rm H}/{\rm cm}^{-3})=-3$, and temperatures of $T=10^{4}$~K. On the other hand, the higher ionisation component has a much lower column density at $\log (N({\HI})/{\rm cm}^{-2})=14$, while at the same time, it has a much higher temperature of $T=10^{5.5}$~K.
    \item We identified H{\sc i}, N{\sc ii}, C{\sc ii}, Si{\sc ii}, N{\sc v} and O{\sc vi} absorption, along with modest O{\sc i} absorption in the quasar spectrum at the redshift of G2, having a total $\log (N({\HI})/{\rm cm}^{-2})=17.33$. This absorption can be divided into six velocity components, spanning a velocity range of 338~km~s$^{-1}$. Five of these components are in a lower ionisation phase, while the other one is in a higher ionisation phase. Three of the low-ionisation clouds have supersolar metallicities, while the metallicities of the other two low-ionisation clouds and the high-ionisation cloud are subsolar. All of the low ionisation clouds have lower temperatures of $10^{4}$~K, compared to the higher temperature of the higher ionisation cloud of $10^{5.5}$~K. The clouds have a wide range of sizes (ranging from $10^{-5}$ to 10~kpc) and column densities (ranging $13 \lesssim\log (N({\HI})/{\rm cm}^{-2})\lesssim17$).
    \item We find that the kinematics of the absorption associated with G2 is more complex than that of G1, as it spans a wider range of velocities and can be divided into more velocity components. G2 also spans a wider range of metallicities and column densities. This is even more interesting as G2 is twice the distance as G1 away from the quasar. Additionally, some G2 absorption is in a higher ionisation state than found in G1. 
    \item In the case of G1, we were able to assign a possible origin for each absorption component to an individual galaxy by comparing the kinematics and metallicities of both the ISM and the CGM. This result suggests that the superposition model from \citet{Bordoloi_2011a} describes this system appropriately. Such an analysis was not as straightforward in the case of G2, indicating that this model does not fit merging galaxies very well. This is probably due to the galaxies being so enmeshed with each other that their gas is also well mixed, compared to G1.
\end{itemize}

In conclusion, this study suggests that the CGM of merging systems is more complex than that of non-interacting galaxies or isolated galaxies. Future work is needed to further disentangle the properties of the CGM in merging galaxies. This work requires a sample of well-defined group environments (e.g., compact groups), along with {\HI} maps to obtain an understanding of star-formation reservoirs in the galaxies themselves.

\section*{Acknowledgements}

We thank Michelle Cluver for the valuable conversations that helped shape this paper in the way it is now. We also thank Enrico Di Teodoro for taking the time to explain how to use 3DBarolo. The authors thank the referee for insightful comments that have improved the manuscript.
Parts of this research were supported by the Australian Research Council Centre of Excellence for All Sky Astrophysics in 3 Dimensions (ASTRO3D), through project number CE170100013.
The data presented herein were obtained at the W. M. Keck Observatory, which is operated as a scientific partnership among the California Institute of Technology, the University of California and the National Aeronautics and Space Administration. The Observatory was made possible by the generous financial support of the W. M. Keck Foundation. Observations were supported by Swinburne Keck program 2022A\_W230. The authors wish to recognise and acknowledge the very significant cultural role and reverence that the summit of Maunakea has always had within the indigenous Hawaiian community. We are most fortunate to have the opportunity to conduct observations from this mountain.


\section*{Data Availability}

The data underlying this paper will be shared following mutually agreeable arrangements with the corresponding authors.
 



\bibliographystyle{mnras}
\bibliography{refs}








\bsp	
\label{lastpage}
\end{document}